\documentclass[preprint,aps,amsmath,nofootinbib,tightenlines,floatfix]{revtex4}
\usepackage{graphicx, epsfig}

\newcommand{\eeq}{\end{equation}}
\newcommand{\beq}{\begin{equation}}
\newcommand{\bea}{\begin{eqnarray}}
\newcommand{\eea}{\end{eqnarray}}

\begin{document}

\setlength{\unitlength}{1mm}

\title{Phenomenology of the Left-Right Twin Higgs Model}
 
\author{Hock-Seng Goh, Shufang Su}
\affiliation{Department of Physics, University of Arizona, Tucson, AZ 85721
}
\begin{abstract}
The twin Higgs mechanism has recently been proposed to solve the little
hierarchy problem.  We study the implementation of the twin Higgs mechanism
in left-right models.  At TeV scale,  heavy quark and gauge bosons 
appear, with rich collider phenomenology.  In addition, there are extra
Higgses, some of which couple to both the Standard Model 
fermion sector and the gauge sector, while
others  couple to the gauge bosons only.
We present the particle spectrum, and study the general features of 
the collider phenomenology of this class of  model 
at the Large Hadron Collider.

\end{abstract}
 
%\pacs{PACS Numbers: }
 
\maketitle

%\newpage

\section{Introduction}
The Higgs mechanism provides a simple and elegant method to explain the 
electroweak symmetry breaking in the Standard Model (SM).
The Higgs boson, however,  is yet to be found. 
While we do not know whether the Higgs boson exists,
unitarity indicates that new physics is very likely to be found at the 
large hadron collider (LHC) \cite{LeeEG}. 

If the electroweak symmetry is broken by the Higgs mechanism, 
the current lower limit on the mass of a scalar SM Higgs comes
from LEP Higgs searches: $m_{h}>114$  GeV \cite{LEPHiggs}. 
Electroweak precision measurements from LEP and SLC 
set an upper bound on the Higgs mass: $m_{h}< 219$ GeV 
at 95\% C.L. \cite{mhlimit}.  
The leading quadratically divergent radiative corrections to the Higgs mass
require the scale of new physics to be around 
TeV scale.   Otherwise, fine tuning in the Higgs potential becomes
severe.  
On the other hand, 
precision measurements constrain the cutoff scale for new physics to be 
likely above 5$-$10 TeV \footnote{There are, however, strong dynamics models that have been
constructed which have a lower cut-off, while being consistent with precision 
measurements \cite{EWmodel}.}, 
leading to a few percent fine tuning in the Higgs potential. 
This is the so called little hierarchy problem or the LEP paradox \cite{LEPparadox}.

Recently, the twin 
Higgs mechanism has been proposed as a solution to the little hierarchy
problem \cite{twinhiggsmirror, twinhiggsleftright, susytwinhiggs}.  
The Higgses emerge as  pseudo-Goldstone bosons once the global symmetry
is spontaneously broken.  Gauge and Yukawa interactions that break  the 
global symmetry give masses to the Higgses, with the leading order being 
quadratically divergent.  When an additional  discrete symmetry is imposed, the leading 
quadratically divergent terms respect the global symmetry.  Thus they do not 
contribute to the Higgs masses.  The resulting Higgs masses obtain logarithmically
divergent contributions.  The Higgs masses are around the electroweak scale when the 
cutoff is around 5$-$10 TeV.  

The twin Higgs mechanism can be implemented
in different ways.  In the mirror twin Higgs 
models~\cite{twinhiggsmirror}, 
a complete copy of the 
SM is introduced, both the gauge interactions and the particle content. 
The discrete symmetry is identified with mirror parity.  
The leading SM contributions to the Higgs masses are canceled by the 
contributions from the mirror particles.
The particles in the mirror world
communicate with the SM particles only via the Higgs particles. 
For the mirror quarks and leptons, they are charged under the mirror gauge 
groups, not the SM ones.  Therefore,  those 
mirror particles can seldom be produced at colliders.  
The Higgs can decay invisibly into mirror bottom quark.  
The 
coupling between the SM Higgs and the mirror bottom quark is suppressed by 
$v/(\sqrt{2}f)$, comparing to the Standard Model $h\bar{b}b$ coupling.  Here $v$ is the Higgs vacuum expectation value (vev) $v=246$ GeV
and $f$ is the symmetry breaking scale in the mirror twin Higgs model,  which is typically around 
800 GeV.  Numerically, the invisible Higgs decay branching ratio  is about $5\%$. 
The searches for invisible Higgs decay at the LHC have been studied in the 
literature \cite{invisible, zeppenfeld, cmsWBF,atlasWBF, atlasZH, atlasttH}.
Analyses in Ref.~\cite{zeppenfeld} show that for a Higgs mass of 120 GeV with SM production cross section, an invisible decay branching ratio of about 13\%
(5\%) can be probed at 95\% C.L. for an integrated luminosity of 10 
(100) ${\rm fb}^{-1}$ at the LHC via weak boson fusion process.
Following the strategy in Ref.~\cite{zeppenfeld},   more detailed analyses including detector simulation at ATLAS \cite{atlasWBF}  show that an invisible Higgs decay branching ratio of about 36\% (25\%) can be probed at 95\% C.L. at the LHC with 10 (30) ${\rm fb}^{-1}$ integrated luminosity.  
More recent analyses \cite{cmsWBF} show that CMS should be able to probe an invisible Higgs decay branching ratio as low as 12\% with 10 ${\rm fb}^{-1}$.
Results based on analyses with $Zh$, $Wh$ \cite{atlasZH} or $t\bar{t}h$ \cite{atlasttH} production channel are less competitive.
Therefore, a measurement of $5\%$ invisible Higgs decay would be possible at the LHC.

The twin Higgs mechanism can also be implemented in left-right models 
with the discrete symmetry being identified with left-right 
symmetry \cite{twinhiggsleftright}.  
In the left-right twin Higgs (LRTH) model, 
the global symmetry is ${\rm U}(4)\times{\rm U}(4) $, with a gauged 
${\rm SU}(2)_L\times {\rm SU}(2)_R \times {\rm U}(1)_{B-L}$ subgroup. 
After Higgses obtain 
vacuum expectation
values, the global symmetry
${\rm U}(4) \times{\rm U}(4)$ breaks down to 
${\rm U}(3)\times{\rm U}(3)$, and 
${\rm SU}(2)_R \times {\rm U}(1)_{B-L}$ breaks down to the SM ${\rm U}(1)_Y$.
Three Goldstone bosons  are eaten by the 
massive gauge bosons $Z_H$ and $W_H^{\pm}$, while the remaining Goldstone 
bosons
contain the SM ${\rm SU}(2)_L$ Higgs doublet and extra Higgses.
The leading quadratically divergent 
SM gauge boson contributions to the Higgs masses are canceled by the 
loop involving the 
heavy gauge bosons.  A vector top 
singlet pair is introduced to generate an ${\cal O}(1)$ top Yukawa coupling. 
The quadratically divergent SM top contributions to the Higgs potential are canceled by 
the contributions from a heavy top partner.  
Many new particles which have
order of one interaction strength with the SM sector are predicted
and rich phenomenology is expected at the LHC.  

This paper is organized as follows. Sec.~\ref{sec:model} describes the 
LRTH
model in detail.  We present the particle content, 
and the structure of gauge and Yukawa interactions.  After spontaneous symmetry breaking, 
we calculate the particle spectrum, and write down the resulting Feynman rules for the 
interactions.  We demonstrate the twin Higgs mechanism in 
Sec.~\ref{sec:twinhiggsmechanism}. 
In Sec.~\ref{sec:spectrum}, we show numerical 
values of the particle masses.  In Sec.~\ref{sec:constraints}, we 
summarize the current experimental constraints on the model parameters.
In Sec.~\ref{sec:signal}, we discuss in detail the collider phenomenology 
of the left-right twin Higgs model.  We analyze the particle production 
cross sections, and their decay patterns.  Sec.~\ref{sec:M0} is devoted to the 
discussion of the case when the mass mixing between the extra vector 
top quark singlet is zero or very small ($\lesssim$ 1 GeV).
The collider signatures are completely different in this limit.
In Sec.\ref{sec:conclusion}, we conclude.  In the appendices, we present
the representation of the Higgs fields in the nonlinear sigma model, the 
exact expressions for the new particle masses, and a complete list of the 
Feynman rules.

\section{The left-right twin Higgs model}
\label{sec:model}
To implement the twin Higgs mechanism we need a global symmetry, which
is partially gauged and spontaneously broken, and a discrete twin
symmetry. In the LRTH model proposed in \cite{twinhiggsleftright},
the global symmetry is ${\rm U}(4)\times {\rm U}(4)$.
The diagonal subgroup of the U(4)$\times$U(4), which is
generated by
\begin{eqnarray}
    \left(%
\begin{array}{cc}
 \frac{1}{2}\sigma_i& 0 \\
  0 & 0 \\
\end{array}%
\right),\left(%
\begin{array}{cc}
  0 & 0 \\
  0 & \frac{1}{2}\sigma_i \\
\end{array}%
\right),\frac{1}{2}\left(%
\begin{array}{cc}
 1 & 0 \\
  0 & 1 \\
\end{array}%
\right)
\label{eq:generator}
\end{eqnarray}
is gauged and identified as the ${\rm SU}(2)_L\times {\rm SU}(2)_R\times
{\rm U}(1)_{B-L}$ gauge group of the left-right model \cite{leftright}. 
Here $\sigma_{1,2,3}$ are three Pauli matrices. 
As explained in Ref.~\cite{twinhiggsleftright}, a bigger
${\rm O}(8)\times {\rm O}(8)$ global symmetry is needed in order to account for
the custodial symmetry at the non-renormalizable level.  However, we
stick to the U(4) language since it makes no significant
difference to the collider phenomenology. The twin symmetry which
is required to control the quadratic divergences is identified
with the left-right symmetry which interchanges L and R. For the
gauge couplings $g_{2L}$ and $g_{2R}$ of
${\rm SU}(2)_L$ and ${\rm SU}(2)_R$, the
left-right symmetry implies that $g_{2L}=g_{2R}=g_2$.

Two Higgs fields, $H$ and $\hat{H}$, are introduced and each
transforms as $(\textbf{4},\textbf{1})$ and
$(\textbf{1},\textbf{4})$ respectively under the global symmetry.
They can be written as
\begin{equation}
H=\left(
\begin{tabular}{c}
$H_L$\\
$H_R$
\end{tabular}
\right),\ \ \ \ \
\hat{H}=\left(
\begin{tabular}{c}
$\hat{H}_L$\\
$\hat{H}_R$
\end{tabular}
\right),
\end{equation}
where $H_{L,R}$ and
$\hat{H}_{L,R}$ are two component objects which are
charged under 
the ${\rm SU}(2)_L\times {\rm SU}(2)_R\times {\rm U}(1)_{B-L}$ as

\begin{equation}
    H_L {\rm \ and\ }\hat{H}_L: ({\bf 2},{\bf 1},1),\ \ \
    H_R {\rm \ and\ }\hat{H}_R: ({\bf 1},{\bf 2},1).
\end{equation}
Each Higgs acquires a non-zero vev as 
\begin{eqnarray}\label{eq:vev1}
    <H> = \left(%
\begin{array}{c}
  0 \\
  0 \\
  0 \\
  f \\
\end{array}%
\right),\;\;\;\;\;
<\hat{H}> = \left(%
\begin{array}{c}
  0 \\
  0 \\
  0 \\
  \hat{f} \\
\end{array}%
\right),
\end{eqnarray}
which breaks one of the U(4) to U(3) and yields seven
Nambu-Goldstone bosons and one massive radial mode. The Higgs
vevs also break ${\rm SU}(2)_R\times {\rm U}(1)_{B-L}$ down to
the SM ${\rm U}(1)_Y$.
The SM hypercharge is given by
\begin{equation}
    \frac{Y}{2} =T_{3R} + \frac{n_{B-L}}{2},
%\nonumber\\
%    Q &=& \frac{1}{2}(\sigma_L^3+Y)
\end{equation}
where $T_{3R}=\sigma_{3R}/2 $ is the third component of ${\rm SU(2)}_R$
isospin, and $n_{B-L}$ is the $B-L$ charge.
We have used the normalization
that the hypercharge of the left handed quarks is $\frac{1}{3}$.
Three Goldstone bosons are eaten by the massive gauge bosons and become their
longitudinal components. The remaining eleven massless Goldstone
bosons are the
SM ${\rm SU(2)}_L$ Higgs doublet from $H_L$, an extra ${\rm
SU(2)}_L$ Higgs doublet from $\hat{H}_L$, a neutral real
pseudoscalar and a pair of charged scalar fields,
which come from the combination of $H_R$ and $\hat{H}_R$ 
\footnote{Once we use the representation of $H$ and $\hat{H}$ in nonlinear sigma model,
small mixtures between Higgses appear, as shown explicitly in 
Eq.~(\ref{GBrep}). }. The gauge
interactions (and Yukawa interactions to be discussed later) break
the global symmetry, which generate a potential for the Goldstone bosons, in
particular, for the SM Higgs doublet.  The left-right discrete
symmetry ensures that the global symmetry is respected at the
quadratic order and so the quadratically divergent mass correction
contributes only to the masses of the already massive radial modes but not to the
masses of the Goldstone bosons. The sub-leading contribution is only
proportional to $\ln\Lambda$, for $\Lambda$ being the cut off
scale.  No severe fine tuning is introduced for $\Lambda$ of the
order of  5$-$10 TeV.

After the Higgses obtain vevs as shown in Eq.~(\ref{eq:vev1}), three
of the four ${\rm SU}(2)_R\times {\rm U}(1)_{B-L}$ gauge bosons become
massive, with masses proportional to $\sqrt{f^2+\hat{f}^2}$. Since
these gauge bosons couple to the SM  matter fields,
their masses are highly constrained from either precision measurements or direct
searches.  Requiring
$\hat{f} \gg {f}$, the masses of the extra gauge bosons can be set
to be large enough to avoid the constraints from the electroweak
precision measurements.  The large value of $\hat{f}$ does not
reintroduce the fine tuning problem for the Higgs potential, since
the gauge boson contributions to the Higgs potential is suppressed
by the smallness of the gauge couplings. 
By imposing certain discrete
symmetry as described below in Sec.~\ref{sec:matter}, the Higgs
field $\hat{H}$ couples only to the gauge sector, 
but not to the SM fermions, in particular, the top quarks.
The top sector only couples to $H$, with a smaller vev $f$.
The top sector
contributions to the Higgs potential, with an unsuppressed ${\cal O}(1)$ top Yukawa
coupling, is therefore under control.

\subsection{Higgs fields in the nonlinear sigma model}
\label{sec:nonlinear} The massive radial modes in $H$ and
$\hat{H}$ obtain masses $ \sim 4\pi f (\hat{f})$  in the
strongly coupled limit. Below the cut off scale $\Lambda$, the
radial modes are integrated out and the effective theory can be
described by a nonlinear sigma model of the 14 Goldstone bosons. 
In our analysis, we
focus on the case where $\Lambda=4 \pi f$.  The results of our studies do not change 
much for $\Lambda=2\pi f$.

The scalar fields can be parameterized by
\begin{equation}
    H = f e^{i\frac{\pi}{f}}\left(%
\begin{array}{c}
  0  \\
  0  \\
  0  \\
  1  \\
\end{array}%
\right),\ \ \ \ \ {\rm with\ }
    \pi = \left(%
\begin{array}{cccc}
  -N/2 & 0 & 0 & h_1 \\
  0 & -N/2 & 0 & h_2 \\
  0 & 0 &  -N/2 & C \\
  h_1^* & h_2^* & C^* & 3N/2  \\
\end{array}%
\right),
\label{eq:Higgsrep}
\end{equation}
where $\pi$ are the corresponding Goldstone fields. $N$ is a neutral real
pseudoscalar, $C$ and $C^*$ are a pair of charged complex scalar
fields, and $h_{\rm SM}=(h_1,h_2)$ is the SM ${\rm SU}(2)_L$ Higgs
doublet. They together comprise the seven Goldstone bosons. $\hat{H}$ can be
parameterized in the same way by its own Goldstone fields $\hat{\pi}$,
which contains $\hat{N}$, $\hat{C}$ and $\hat{h}=(\hat{h}_1^+,
\hat{h}_2^0)$.

When symmetry is further broken by the vev of $h_{\rm SM}$:
$\langle h_{\rm SM}\rangle =(0,v/\sqrt{2})$, 
electroweak symmetry ${\rm SU}(2)_L\times {\rm U}(1)_Y$
is broken down
to ${\rm U}(1)_{\rm EM}$. On the other hand, 
$\hat{h}$ does not get a vev. 
We can rewrite the two steps of symmetry breaking in
one single step, with the vevs
of $H$ and $\hat{H}$ being
\begin{equation}
\label{eq:vev2}
    <H> =\left(%
\begin{array}{c}
  0\\
  if\sin x \\
  0\\
  f\cos x\\
\end{array}%
\right),\ \ \ \ \
    <\hat{H}> = \left(%
\begin{array}{c}
0\\
  0 \\
  0\\
\hat{f}\\
\end{array}%
\right),
\end{equation}
where $x=\frac{v}{\sqrt{2}f}$. The original gauge symmetry ${\rm
SU}(2)_L\times{\rm SU}(2)_R\times{\rm U}(1)_{B-L}$ is broken down
to ${\rm U}(1)_{\rm EM}$ and generates four charged and two neutral gauge bosons
: $W^{\pm}$, $W^{\pm}_H$, $Z$ and $Z_H$. $W$ and $Z$ are the usual
massive gauge bosons in the SM and $W_H$, and $Z_H$ are three
additional massive gauge bosons with masses of a few TeV. Six out
of the fourteen Goldstone bosons are eaten by the massive gauge bosons. By studying
the charges of the Goldstone fields and the symmetry breaking pattern,
we know that $h_1$ and the imaginary component of $h_2$ are eaten by
$W$ and $Z$, as in the SM case. One linear combination of $C$ and
$\hat{C}$ and one linear combination of $N$ and $\hat{N}$ are
eaten by $W_H$ and  $Z_H$, respectively. To simplify our analysis,
we work in the unitary gauge so that all the fields that are eaten by
the massive gauge bosons are absent in the following discussions.  After the
re-parametrization of the fields, with the details to be found in
Appendix~\ref{app:unitary}, we are left with one neutral pseudoscalar 
$\phi^0$, a pair of charged scalar $\phi^\pm$, the SM
physical Higgs $h$, and a ${\rm SU}(2)_L$ doublet
$\hat{h}=(\hat{h}_1^+, \hat{h}_2^0)$.

In general,  the interactions among the various particles do not respect the global
symmetry and are only required to be gauge invariant. Therefore,
we use the representations of ${\rm SU}(2)_L\times {\rm SU}(2)_R$ instead of
${\rm SU}(4)$ when writing down the interactions. The easiest way to
write down the leading gauge invariant interactions involving the Goldstone bosons is
to begin with the linear fields and 
set all the radial modes to zero.  We therefore write down the
linear model as given in \cite{twinhiggsleftright} and replace $H$ and
$\hat{H}$ by their nonlinear expressions given in Eqs.~(\ref{GBrep})
and (\ref{unitary}).

The Lagrangian can be written as
\begin{eqnarray}\label{lagrangian}
    {\cal L} = {\cal L}_H+{\cal L}_G+{\cal L}_f+{\cal L}_Y+{\cal L}_{one-loop}
+{\cal L}_{\mu}.
\end{eqnarray}
The various pieces in Eq.~(\ref{lagrangian}), in the order in which they are
written, are covariant kinetic terms for Higgses, gauge
bosons and fermions, Yukawa interactions, one-loop Coleman-Weinberg (CW)
potential~\cite{CW} for Higgses and soft symmetry breaking $\mu$
terms.

Once $H$ and $\hat{H}$ obtain vevs, the Higgs kinetic term ${\cal
L}_H$ gives rise to the gauge boson mass terms.  Using the
nonlinear Higgs representation given in Eq.~(\ref{GBrep}) and the
unitary gauge choice given in Eq.~(\ref{unitary}), we 
obtain the derivative self-interactions of the scalars and the interactions
between scalars and gauge bosons. 
The kinetic term for the gauge bosons,
${\cal L}_G$, is standard.  It  gives us three and four gauge
boson  self-couplings. 
The covariant kinetic term for
fermions, ${\cal L}_f$, is straight forward to write down once the
gauge representations of all fermions are known.  It gives rise to the gauge
interactions of fermions. 
The Yukawa coupling ${\cal L}_Y$ couples fermions
to Higgses.  It generates the fermion masses once Higgses get
vevs.  It also gives rise to scalar-fermion-fermion Yukawa
interactions. U(4) violating interactions, i.e. the gauge
couplings and Yukawa couplings, generate a potential for the Goldstone bosons at
loop level, which is indicated by ${\cal L}_{one-loop}$ for the
one-loop contribution. In particular, it generates mass terms for
the Goldstone Higgses. 
%These couplings, however, are required to
%preserve the discrete symmetry $L\leftrightarrow R$ (Twin
%symmetry) in order to protect the pseudo Goldstone bosons from
%getting quadratically divergent masses. While it is possible to
%make the top Yukawa invariant under the U(4) global symmetry, we
%will consider only the minimal model in Ref.~\cite{twinhiggsleftright} where
%the top Yukawa interaction preserve only the left-right symmetry.
The neutral scalar $\phi^0$, however, remains massless due to a
residual U(1) global symmetry.
A `$\mu$-term' is introduced to break the global
 $U(1)$ symmetry softly in order to give a
mass to $\phi^0$. This $\mu$-term inevitably gives 
masses to other scalars.  Other $\mu$-terms could be added to 
generate masses for other Higgses, for example, the dark matter candidate $\hat{h}_2^0$.

In the following subsections, we discuss in detail each
individual term in the Lagrangian, and obtain the particle
spectrum and interactions.

\subsection{Gauge bosons}
\label{sec:gauge}
Given the generators
of ${\rm SU}(2)_L\times {\rm SU}(2)_R\times {\rm U}(1)_{B-L}$ as
shown in Eq.~(\ref{eq:generator}), the corresponding gauge fields
are
\begin{equation}
    W_2 =\frac{1}{2}\left(%
\begin{array}{cccc}
  W^0_L & \sqrt{2}W^+_L & 0 & 0 \\
  \sqrt{2}W^-_L & -W^0_L & 0 & 0 \\
   0 & 0 & W^0_R & \sqrt{2}W^+_R \\
   0 & 0 & \sqrt{2}W^-_R & -W^0_R \\
\end{array}%
\right),\ \ \
    W_{B-L} =\frac{W_1}{2}
\left(%
\begin{array}{cccc}
  1 & 0& 0 & 0 \\
 0 & 1 & 0 & 0 \\
   0 & 0 & 1& 0 \\
   0 & 0 & 0 & 1 \\
\end{array}%
\right),
\end{equation}
where for simplicity, we have suppressed Lorentz indices. $W_2$ contains the gauge
fields $(W_L^\pm,W_L^0)$ for ${\rm SU}(2)_L$ and $(W_R^\pm,W_R^0)$
for ${\rm SU}(2)_R$ , and $W_1$ is the gauge field corresponding
to $U(1)_{B-L}$. The covariant derivative is
\begin{equation}
 D^{\mu} = \partial^{\mu} - i g_2 W_2^{\mu} -ig_1n_{B-L}W_{B-L}^{\mu},
\end{equation}
where $g_1$ and $g_2$ are the gauge couplings for ${\rm U}(1)_{B-L}$ and
${\rm SU}(2)_{L,R}$, and
$n_{B-L}$ is the charge of the field under ${\rm U}(1)_{B-L}$.

The covariant kinetic terms of Higgs fields can
be written down as
\begin{eqnarray}
    {\cal L}_H &=& (D_{\mu}H)^{\dagger}D^{\mu}H +
(D_{\mu}\hat{H})^{\dagger}D^{\mu}\hat{H}, 
\end{eqnarray}
with $n_{B-L}=1$. When $H$ and $\hat{H}$ get vevs as shown in 
Eq.~(\ref{eq:vev2}), ${\rm SU}(2)_L\times {\rm SU}(2)_R\times {\rm
U}(1)_{B-L}$ breaks down to ${\rm U}(1)_{\rm EM}$.  There are six
massive gauge bosons $W^{\pm}$, $W_H^{\pm}$, $Z$, $Z_H$, and one
massless photon $\gamma$. For the charged gauge bosons,  there is
no mixing between the $W_L^\pm$ and the $W_R^\pm$: $W^{\pm}=W_L^{\pm}$
and $W_H^{\pm}=W_R^{\pm}$. Their masses are
\begin{equation}
    m^2_{W} = \frac{1}{2}g_2^2 f^2 \sin^2x,\ \ \
    m^2_{W_H} = \frac{1}{2}g_2^2 F^2,
\end{equation}
where $F^2=\hat{f}^2+f^2\cos^2x$. The neutral gauge bosons $Z_H$,
$Z$ and $\gamma$ are linear combinations of $W_L^0$, $W_R^0$ and
$W_1$:
\begin{equation}\label{eq:mixing}
    \left(%
\begin{array}{c}
  Z_H \\
  Z \\
  \gamma \\
\end{array}%
\right) = U \left(%
\begin{array}{c}
  W_R^0 \\
  W_L^0 \\
  W_1 \\
\end{array}%
\right),\ \ \ {\rm where}\
    U \sim \left(%
\begin{array}{ccc}
  \frac{\sqrt{\cos2\theta_w}}{\cos\theta_w} & \frac{\sqrt{\cos2\theta_w}\sin^2\theta_w}{\cos^3\theta_w}\frac{m_W^2}{m_{W_H}^2} & -\frac{\sin\theta_w}{\cos\theta_w} \\
  -\frac{\sin^2\theta_w}{\cos\theta_w} & \cos\theta_w & -\frac{\sin\theta_w\sqrt{\cos2\theta_w}}{\cos\theta_w} \\
  \sin\theta_w& \sin\theta_w & \sqrt{\cos2\theta_w} \\
\end{array}%
\right),
\end{equation}
to the leading order in $v/f$, and $\theta_w$ is the Weinberg angle. 
We see that $Z_H$ is mainly a
linear combination of $W_R^0$ and $W_1$.  A small component of
$W_L^0$ in $Z_H$, which is  of the order of $v^2/f^2$, 
appears after electroweak symmetry breaking. 
For $Z$ and $\gamma$, all three
of $W_L^0$, $W_R^0$ and $W_1$ contribute at leading order.  This
is because the hypercharge gauge boson $B$ is a linear combination
of $W_R^0$ and $W_1$, while $Z$ and $\gamma$ are linear
combinations of $B$ and $W_L^0$. The masses of $Z$ (at leading
order in $v/f$) and $Z_H$ are
\begin{eqnarray}
m_{Z}^2 &\sim &
\frac{g_2^2+g_Y^2}{g_2^2}m^2_{W}\left[1-(\frac{g_Y}{g_2})^4\frac{m^2_{W}}{m^2_{W_H}}\right], \\
 m_{Z_H}^2 &=&\frac{g_1^2+g_2^2}{g_2^2}(m^2_{W_H}+m^2_{W})-m_{Z}^2,
\end{eqnarray}
where $g_Y$ is the usual hypercharge coupling in the SM as given below in 
Eq.~(\ref{eq:coupling}).
The exact expression for the mixing matrix $U$ and
the gauge boson mass eigenvalues can be found in Appendix~\ref{app:masses}.
The gauge couplings $g_1$, $g_2$, and $g_Y$ are 
related to $e$ and Weinberg angle $\theta_w$ as
\begin{equation}
    g_1=\frac{e}{\sqrt{\cos2\theta_w}},\ \ \
    g_2=\frac{e}{\sin\theta_w},\ \ \ 
    g_Y=\frac{e}{\cos\theta_w}.
\label{eq:coupling}
\end{equation}

The gauge boson kinetic term ${\cal L}_G$
is similar to that of the SM,  with an
exact copy for the right handed gauge bosons:
\begin{equation}\label{eq:gaugekinetic}
{\cal L}_G=-\frac{1}{2}{\rm tr}(F_{\mu\nu})_L(F^{\mu\nu})_L-
\frac{1}{2}{\rm tr}(F_{\mu\nu})_R(F^{\mu\nu})_R- \frac{1}{4}{\rm
tr}(F_{\mu\nu})_{B-L}(F^{\mu\nu})_{B-L},
\end{equation}
where $(F_{\mu\nu})_{L,R}$ and  $(F_{\mu\nu})_{B-L}$ are the field
strength for ${\rm SU}(2)_{L,R}$ and ${\rm U}(1)_{B-L}$,
respectively. With the help of the transformation matrix $U$ given
above, self-couplings between gauge boson mass eigenstates can be
derived. We summarize these interactions in
Table~\ref{tab:gaugeself} in
Appendix~\ref{app:LHinteraction}.

\subsection{Matter sector}
\label{sec:matter}

The SM quarks and leptons (with the addition of three right-handed
neutrinos) are charged under ${\rm SU}(3)_c\times {\rm
SU}(2)_L\times {\rm SU}(2)_R\times {\rm U}(1)_{B-L}$ as
\begin{eqnarray}
    L_{L\alpha}=-i\left(\begin{array}{c}~\nu_{L\alpha}
\\l_{L\alpha}\end{array}\right):
({\bf 1},{\bf 2},{\bf 1},-1), \ \ \ \ \ &&
    L_{R\alpha}=\left(\begin{array}{c}~\nu_{R\alpha}
\\l_{R\alpha}\end{array}\right): 
({\bf 1},{\bf 1},{\bf 2},-1),\nonumber\\
    Q_{L\alpha}=-i\left(\begin{array}{c} u_{L\alpha}
\\ d_{L\alpha}\end{array}\right): 
({\bf 3},{\bf 2},{\bf 1},1/3),\ \ \ \ \ &&
    Q_{R\alpha}=\left(\begin{array}{c}u_{R\alpha}
\\d_{R\alpha}\end{array}\right):({\bf 3},{\bf 1},{\bf 2},1/3),
\end{eqnarray}
where  ``$\alpha$" is the family index which runs from 1 to 3. The
additional ``$-i$" in the definition of $Q_{L\alpha}$ and
$L_{L\alpha}$ is introduced to make the fermion mass real,
given the Yukawa interactions in Eqs.~(\ref{eq:Yukawa1})
and (\ref{eq:topyukawa}) below. Notice that the SM 
${\rm SU}(2)_L$ singlets $u_{R\alpha}$ and $d_{R\alpha}$
are now grouped together as doublets under ${\rm SU}(2)_R$. Three
generations of right-handed neutrinos $\nu_{R\alpha}$ are
introduced, which combined with $l_{R\alpha}$ to form ${\rm
SU}(2)_R$ doublets. 
%We choose to work with 4 component Dirac
%fermion so the right handed component of $Q_L$ and the left handed
%component of $Q_R$ vanish and similarly for all other fermions.

The masses of the first two generation quarks and bottom quark are
obtained from the  non-renormalizable operators
\begin{equation}
    \frac{y_u^{\alpha\beta}}
{\Lambda}(\bar{Q}_{L\alpha}\tau_2 H_L^*)(H_R^T\tau_2{Q}_{R\beta})
+\frac{y_d^{\alpha\beta}}{\Lambda}(\bar{Q}_{L\alpha}
H_L)(H_R^{\dagger}{Q}_{R\beta})
+ h.c.,
 %   y_{\nu}^{ij}(\bar{L}_{L,i}\tau_2 H_L^*)(H_R^T\tau_2{L}_{Rj})+y_l^{ij}(\bar{L}_{Li} H_L)(H_R^{\dagger}{L}_{Rj}) +
%h.c.
\label{eq:Yukawa1}
\end{equation}
where $\tau_2=\left(%
\begin{array}{cc}
  0 & -1\\
  1 & 0\\
\end{array}%
\right)$. Once $H_R$ obtains a vev, it generates effective Yukawa couplings
for the quarks of the order of $f/\Lambda$.  Similar terms can
be written down for the lepton sector, which generate small
masses for the charged leptons, and Dirac mass terms for
the neutrinos.   In addition, 
we can write down an operator 
$({L}_R^T \tau_2 \hat{H}_R)C( \hat{H}_R^T \tau_2 L_R)/\Lambda$,  with 
$C$ being the charge conjugation operator.  Such term generates 
large Majorana  masses of the order of $\hat{f}^2/\Lambda$ for $\nu_R$.
The smallness of the usual neutrino masses can be achieved via the 
seesaw mechanism.  

Such non-renormalizable  operators, with effective Yukawa
couplings suppressed by $f/\Lambda$, cannot account for the ${\cal O}(1)$ top
Yukawa. In order to give the top quark a mass of the order of electroweak scale, 
a pair of vector-like quarks
\begin{equation}
    q_L: ({\bf 3},{\bf 1},{\bf 1},4/3),\ \ \ \ \
    q_R: ({\bf 3},{\bf 1},{\bf 1},4/3),
\end{equation}
are introduced, which are singlets under ${\rm SU}(2)_L\times{\rm
SU}(2)_R$. The gauge invariant top Yukawa terms can then be written
down as
\begin{equation}
    y_L\bar{Q}_{L3}\tau_2 H_L^*q_R+y_R\bar{Q}_{R3}\tau_2H_R^*q_L - M\bar{q}_Lq_R + h.c.
\label{eq:topyukawa}
\end{equation}
where $Q_{L3} = -i(u_{L3},d_{L3})$ and $Q_{R3} = (u_{R3},d_{R3})$.
%Again, the additional``$i$" in the definition of $Q_{L3}$ is just
%to make the fermion mass real.  
Under left-right symmetry, $y_L=y_R=y$.
Once Higgses $H_{L,R}$ get  vevs,
the first two terms in Eq.~(\ref{eq:topyukawa}) generate masses
for a SM-like top quark $(u_{L3},q_R)$ with mass $yv/\sqrt{2}$, and
a heavy top quark $(q_L, u_{R3})$ with mass $yf$. In
Eq.~(\ref{eq:topyukawa}), we also include the mass mixing term
$M\bar{q}_Lq_R$, which is allowed by gauge invariance.
A non-zero value of $M$ leads to the mixing between the SM-like top
quark and the heavy top quark. The mass eigenstates, heavy top
$T$ and light top $t$, are mixtures of the gauge eigenstates:
\begin{equation}\label{eq:topmix}
    \left(%
\begin{array}{c}
  T_L \\
  t_{L} \\
\end{array}%
\right) =\left(%
\begin{array}{cc}
  \cos \alpha_L & \sin \alpha_L\\
  -\sin \alpha_L& \cos \alpha_L\\
\end{array}%
\right)\left(%
\begin{array}{c}
  q_L \\
  u_{L3} \\
\end{array}%
\right),\ \ \
\left(%
\begin{array}{c}
  T_R \\
  t_{R} \\
\end{array}%
\right) = \left(%
\begin{array}{cc}
  \cos \alpha_R & \sin \alpha_R\\
  -\sin \alpha_R& \cos \alpha_R\\
\end{array}%
\right)\left(%
\begin{array}{c}
  u_{R3} \\
  q_R \\
\end{array}%
\right),
\end{equation}
with the mixing angles $\alpha_L$ and $\alpha_R$ for the left- and
right-handed fields.  The larger the value of $M$, the larger the
mixing between the two gauge eigenstates.  In particular, the
left-handed light top quark has a non-negligible component of ${\rm
SU}(2)_L$ singlet $q_L$ once $M$ is large. The value of $M$ is
constrained by the requirement that the branching ratio of $Z\rightarrow b \bar{b}$
remains consistent with the experiments.
It is also constrained by the 
oblique parameters.
%The deviation of $Z\bar{b}b$ from its SM predictions, introduced
%by top quarks running in the loop,  is tightly
%constrained from  $Z$-pole precision measurements.
In  our analysis, we took $M$ to be small, and picked a typical
value of $M$=150 GeV. Our results do not change much if other
small values of $M$ is used.  However, once $M$ is very small $\lesssim$ 
1 GeV, or in the limit that $M=0$, the collider phenomenology changes 
significantly, which will be discussed in Sec.~\ref{sec:M0}.

The masses of the light SM-like top and  the heavy top are
\begin{eqnarray}
    m_t^2 &\sim&  y^2f^2\sin^2 x-M^2\sin^2 x \sim (yv/\sqrt{2})^2,\\
    m_T^2 &=&  y^2f^2+M^2-m_t^2.
\end{eqnarray}
The top Yukawa coupling can then be determined by fitting the
experimental value of the light top quark mass. The top quark 
mixing angles can be written 
in term of these physical masses. At the leading
order of $M/f$ and  $\sin x$, the mixing angles are
\begin{equation}\label{topmixing}
    \sin \alpha_L\sim \frac{M}{m_T}\sin x,\ \ \
    \sin \alpha_R\sim \frac{M}{m_T}(1+ \sin^2x),
\end{equation}
which are usually small.
%The mixing angles $\alpha_L$ and $\alpha_R$ are considered to be
%close to zero. 
For the SM-like light top quark $t$, the left-handed component
is mostly $u_{L3}$, while the right-handed component is mostly
$q_R$. 
%This is different from the usual SM top quark, where the
%right-handed component is $u_{R3}$.
This is different from the
first two generations where the right handed components of the
up-type quark are $u_{R\alpha}$ that couple to $W_H^{\pm}$. 
%In particular, $Z\bar{t}_Rt_R$ in LRTH differs from its
%value of the Standard Model.  The current experiments, however,
%cannot provide any constraints since only strong interaction
%$G\bar{t}t$ and ${\rm SU}(2)_L$ coupling $W\bar{t}_L b_L$  can be
%probed.  The former is the same in the LRTH and the SM. 
%For $W\bar{t}_L b_L$, the precision of the current 
%measurement is not sensitive enough
%to probe the small deviation, which is of the order of
%$\frac{M}{m_T}\frac{v}{f}$. 
The right-handed component of $t$ in
the LRTH model is also different from the little Higgs models
\cite{lhtao}, where $t_R$ is an
${\cal{O}}(1)$  mixture of $u_{R3}$ and $q_R$. 
While it is difficult to distinguish the light top quark in the LRTH from the SM top quark, or 
the light top quark in the little Higgs models at 
current colliders, future measurements at the LHC of the right-handed coupling of $t$ to the heavy 
gauge bosons could provide important clue. The exact formulas
for the mixing angles and the mass eigenvalues for the SM-like top quark
and the heavy top quark can be found in
Appendix~\ref{app:masses}.

In principle, we could  also write down similar Yukawa terms for the other scalar
field $\hat{H}$.  However, this would give the heavy top quark a much larger
mass of the order of $y\hat{f}$. Such a large value of the heavy top
quark mass reintroduces the fine tuning problem in the Higgs potential
since the top quark induced loop correction is too large.
To avoid this, a parity is introduced in
the model under which $\hat{H}$ is odd while all the other fields are
even. This parity thus forbids renormalizable coupling between
$\hat{H}$ and fermions, especially the top quark. Therefore,
at renormalizable level, 
$\hat{H}$ couples only to the gauge boson sector, while $H$
couples to both the gauge sector and the matter fields. The lightest
particle that is odd under this parity, the neutral
$\hat{h}_2^0$, is stable, and therefore constitutes  a good dark matter candidate.

The interactions between the Higgses and 
top quarks, can be obtained from
Eq.~(\ref{eq:topyukawa}) once the top quarks are rotated into
their mass eigenstates.  The Yukawa interactions of the other fermions can be obtained from 
Eq.~(\ref{eq:Yukawa1}), which is proportional to the fermion masses.
The Feynman rules can be found in
Table~\ref{tab:higgsfermion} in Appendix~\ref{app:LHinteraction}.
%We have ignored the couplings between the Higgses and the 
%first two generations due to their small masses.

The Lagrangian for the fermion kinetic term can be written down as
\begin{eqnarray}
{\cal{L}}_f &=& \bar{L}_{\alpha}i\gamma_{\mu}(\partial^{\mu}
-ig_2W_2^{\mu}+ig_1W_{B-L}^{\mu})L_{\alpha}+
\bar{Q}_{\alpha}i\gamma_{\mu}(\partial^{\mu}
-ig_2W_2^{\mu}-i\frac{g_1}{3}W_{B-L}^{\mu})Q_{\alpha}\nonumber\\
&+&\bar{q}i\gamma_{\mu}(\partial^{\mu}
-i\frac{4g_1}{3}W_{B-L}^{\mu})q
\end{eqnarray}
where we have used $L=(L_L,L_R)$, $Q=(Q_L,Q_R)$ and $q=(q_L, q_R)$.
We have ignored the strong interactions for the quarks, which are 
the same as those in the SM.
The fermion gauge interactions can be found in 
Table~\ref{tab:gaugefermion} in Appendix~\ref{app:LHinteraction}.

It is worth noting that the mixing angles between the light top quarks
and the heavy ones are proportional to $M$. 
In the special case when $M$ is set to
zero,  there is no mixing between the two. 
The light top quark $t$ is made purely
of $(u_{L3}, q_R)$, while $T$ is made purely of $(q_L, u_{R3})$. 
Certain couplings go to zero in this limit.
\begin{itemize}
\item{gauge couplings:} 
For $M=0$, 
$W_H$ only couples to $\bar{T}b$, but not $\bar{t}b$.
This is because both
the left- and right-handed components of the light top are singlet
of ${\rm SU}(2)_R$. Therefore, the ${\rm SU}(2)_R$ weak
interactions of the light top quark do not exist.  
Similarly, $W$ only couples to $\bar{t}b$, but not $\bar{T}b$.  
For $Z$ and $Z_H$, they 
only couple to $\bar{t}t$ and $\bar{T}T$, but not the mixture of these two. 
\item{Top Yukawa couplings:} 
It is obvious from Eq.~(\ref{eq:topyukawa}) that $\phi^\pm$
and $\phi^0$,
which reside in $H_R$, at renormalizable level couple only to
$T$ if $M=0$. While the SM Higgs $h$, which resides 
in $H_L$, couples only to the light top quark $t$ at the order of $(v/f)^0$.
A small $h \bar{T}T$ coupling
(suppressed by $v/f$) appears after the nonlinear Higgs fields are expanded to 
higher orders.
\end{itemize}
In Table~\ref{tab:M0}, we summarize the non-vanishing and vanishing 
gauge and third generation Yukawa couplings in the $M=0$ limit.
The vanishing of those couplings leads to dramatic changes in the
collider phenomenology, which will be discussed in
Sec.~\ref{sec:M0} below.

\begin{table}
\begin{tabular}{|l|c|c|}\hline
&Non-vanishing couplings&Vanishing couplings \\ \hline
Gauge couplings&$W\bar{t}b$,$W_H\bar{T}b$, $Z\bar{T}T$,$Z\bar{t}t$
&$W\bar{T}b$, $W_H\bar{t}b$, $Z\bar{T}t$, $Z_H\bar{T}t$\\
Yukawa couplings&$\phi^0\bar{T}T$, $\phi^0\bar{b}b$, 
$\phi^+\bar{T}b$, $h\bar{t}t$ &
$\phi^0\bar{T}t$, $\phi^0\bar{t}t$, $\phi^{+}\bar{t}b$,
$h\bar{T}t$,  $h\bar{T}T$ \\ \hline
\end{tabular}
\caption{Non-vanishing and vanishing gauge and third generation
Yukawa couplings at the order of $(v/f)^0$ in the $M=0$ limit.}
\label{tab:M0}
\end{table}

\subsection{One-loop Higgs potential}
\label{sec:Loneloop}

The Goldstone bosons $\pi$ and $\hat{\pi}$ are massless at tree level but obtain
masses from quantum effects. The one-loop CW potential is given by \cite{CW}
\begin{equation}
    V = \sum_i \frac{1}{64\pi^2}M_i^4(\ln \frac{M_i^2}{\Lambda^2}+\alpha),
\end{equation}
where the  formula sums over all the particles in the model.
Here $M_i^2$ is the field dependent
squared mass.
The expression for $M_i^2$ for gauge bosons and light/heavy top quarks
can be found in
Appendix~\ref{app:masses}. The constant $\alpha$ is taken to
be $-3/2$. Expanding the potential with respect to the physical
Higgses, we obtain the SM Higgs potential $V_0(h)$, which
determines the SM Higgs vev and its mass, as well as the masses
for the other Higgses $\phi^\pm$, $\phi^0$, $\hat{h}_1^\pm$ and
$\hat{h}_2^0$. The exact expressions for
the Higgs masses can  be found in
Appendix~\ref{app:masses}. 
At leading 
order \footnote{Here and later in the paper when the phrase ``leading order" is used, we mean that the leading order contributions to the interactions or masses are kept.  
%We use the exact expression for the non-linear Higgs field as in Eq.~(\ref{GBrep}) to 
%obtain the vertex without any approximation.  
The expansions of the non-linear Higgs fields are performed up to the fifth order in our analyses.}, 
the Higgs
masses are:
\begin{eqnarray}
m_{h}^2&\sim& \frac{y^4}{2\pi^2}f^2\sin^2x 
(\ln\frac{\Lambda^8}{m_t^3m_T^5}+\frac{13}{4}),\\
    m_{\phi^\pm}^2 &\sim & \frac{3}{16\pi^2}
g_1^2m_{W_H}^2(\ln\frac{\Lambda^2}{m_{Z_H}^2}+1),\\
    {m}_{\hat{h}_2}^2 & \sim & \frac{3}{16\pi^2}\frac{m_{W_H}^2}{2}\left[
    g_2^2(\ln\frac{\Lambda^2}{m_{W_H}^2}+1)+\frac{2g_1^2+g_2^2}{2}(\ln\frac{\Lambda^2}{m_{Z_H}^2}+1)\right],\\
{m}_{\hat{h}_1}^2
&\sim&{m}_{\hat{h}_2}^2+\frac{3}{16\pi^2}
g_1^2m_W^2(\frac{m_{W_H}^2}{m_{Z_H}^2}
\ln\frac{m_{Z_H}^2}{m_{Z}^2}+\ln\frac{\Lambda^2}{m_{Z_H}^2}+1),
\end{eqnarray}
where $m_{W,W_H,Z,Z_H}$ are the gauge boson masses and $m_{t,T}$ are the light top
quark mass and the heavy top mass respectively.  As we now explain, $\phi^0$  remains
massless as it is a Goldstone boson corresponding to a residual  global
${\rm U}(1)$.  The LRTH model has  a
${\rm U}(1)_R\times {\rm U}(1)_{\hat {R}}$ global symmetry where
${\rm U}(1)_R$ transforms only $H_R$ and ${\rm U}(1)_{\hat {R}}$ transforms
only $\hat{H}_R$.  One linear combination of these ${\rm U}(1)$'s
is gauged and the corresponding Goldstone boson becomes the longitudinal 
mode of the massive gauge boson after the symmetry is spontaneously broken.
The orthogonal combination is an exact global
symmetry which is preserved by all interactions.   Therefore, the corresponding Goldstone boson 
$\phi^0$ remains massless even after 
spontaneous symmetry breaking.
A massless neutral
scalar with unsuppressed couplings to the SM fermions and gauge
bosons has already been excluded experimentally. To give mass to
$\phi^0$, we need to introduce a $\mu$-term, as discussed in the
next section.

%Both $U(1)$ are spontaneously broken and two NGbs are
%generated. One of them is eaten by the heavy gauge field $Z_H$ and
%the other one is a real NGb $\phi^0$. 

%Including the $\mu$-terms, at the leading order, the mass spectrum
%of the Higgs sector is
%\begin{eqnarray}
%    m_{\phi}^2 &\sim & \mu_r^2\frac{\hat{f}}{f}\nonumber\\
%    m_C^2 &\sim & \frac{3}{16\pi^2}g_1^2m_{wH}^2(ln\frac{\Lambda^2}{m_{zH}^2}+1)+\mu_r^2\frac{\hat{f}}{f}\nonumber\\
%    \hat{m}_{h_2}^2  &\sim & \frac{3}{16\pi^2}y^4f^2(ln\frac{\Lambda^2}{m_{tH}^2}+1)+\hat{\mu}^2+\mu_r^2\frac{f}{\hat{f}}\nonumber\\
%    \hat{m}_{h_1}^2
%&\sim&\hat{m}_{h_2}^2+\frac{3}{16\pi^2}g_1^2m_w^2(\frac{m_{wH}^2}{m_{zH}^2}ln\frac{m_{zH}^2}{m_{z}^2}+ln\frac{\Lambda^2}{m_{zH}^2}+1)
%\end{eqnarray}

\subsection{$\mu$ terms}
\label{sec:muterm} The following $\mu$-term can be introduced in
the potential:
\begin{equation}\label{eq:muterm}
   V=-\mu_l^2(H_L^{\dagger}\hat{H}_L+h.c.)-\mu_r^2(H_R^{\dagger}\hat{H}_R+h.c.)+ \hat{\mu}^2\hat{H}_L^{\dagger}\hat{H}_L.
\end{equation}
The first term introduces a mixing between $H_L$ and $\hat{H}_L$,
which breaks the parity that we introduced in
sec.~\ref{sec:matter} to forbids the Yukawa coupling between $\hat{H}$ and fermions.
To preserve the stability of $\hat{h}_2^0$
dark matter, we choose $\mu_l= 0$. The second term breaks the U(1)
global symmetry that protects the mass of $\phi^0$ and thus
generates a mass for $\phi^0$. The nonequality between $\mu_l$
and $\mu_r$ breaks the left-right parity, albeit, only softly.
Therefore, it is natural for $\mu_r$ to be of the order of $f$ or
smaller.
$\mu_r$ term also contributes a tree level mass to the SM Higgs:
\begin{eqnarray}
    m_{h}^2 &\sim & \mu_r^2\frac{\hat{f}}{2f}.
\end{eqnarray}
In order not to reintroduce fine tuning, $\mu_r$ has to be less
than about $\frac{f}{4\pi}$. 
In our analysis below, we choose $\mu_r$ to be fairly small,
but   enough to push up $m_{\phi^0}$ above the current experimental
bounds. 

The masses for $\hat{h}_1$ and $\hat{h}_2$, which are
relevant for the dark matter relic density analysis, can be
obtained from one-loop CW potential as explained in
Sec.~\ref{sec:Loneloop}. They are of the order of 200 $-$ 700 
GeV and depend on the Higgs vev $\hat{f}$. Adding the third $\hat\mu^2$
term in Eq.~(\ref{eq:muterm}) allows us to vary the mass of the dark matter 
independently as a free
parameter. Such a term also breaks the left-right symmetry softly.
Therefore, it is natural for it not to be much bigger than $f$.
The masses for the Higgses that are introduced by these $\mu$
terms are:
\begin{eqnarray}
    m_{\phi^0}^2 &\sim & m_{\phi^\pm}^2 \sim \mu_r^2\frac{\hat{f}}{f},\\
    m_{\hat{h}_1}^2 &\sim & m_{\hat{h}_2}^2 \sim
    \mu_r^2\frac{f}{\hat{f}}+\hat{\mu}^2.
\end{eqnarray}

\section{The twin Higgs mechanism in the LRTH model}
\label{sec:twinhiggsmechanism}
In this section, we demonstrate the twin Higgs mechanism explicitly 
in the LRTH model.
The cancellation of the quadratically divergent mass terms of the
pseudo Goldstone bosons 
can be understood in two different ways. The simplest way to
understand the cancellation is by looking at the linear model
which has a ${\rm SU}(2)_L\times {\rm SU}(2)_R$ gauge symmetry.
The most general gauge invariant quadratic terms that can be
written down are

\begin{equation}
    \eta_L H_L^{\dagger}H_L + \eta_R H_R^{\dagger}H_R,
\end{equation}
where $\eta_L$ and $\eta_R$ depend on the particles running in the loop.
Parity  symmetry requires $\eta_L=\eta_R=\eta$ and so the two
terms above can be combined to form a term 
$\eta H^\dagger H$, which is U(4)
invariant.  Since only terms that explicitly break the 
global U(4) symmetry give mass to the Goldstone bosons, 
such quadratic terms do not contribute to the potential
of the Goldstone bosons.
%This show that the quadratic term of linear fields does
%not have contribution to the potential of the pseudo Goldstone
%fields. 
This argument does not depend on the form of
$\eta$'s.  Therefore,  not only the one loop quadratically divergent term
cancel, any quadratic term (finite or logarithmically divergent)
generated at any loop order in perturbation theory is canceled if
the left-right symmetry is exact. On the other hand, if the
left-right symmetry is broken softly by $m^2$, $\eta_L$ and
$\eta_R$ do not have to be the same. However, the difference,
$\eta_L-\eta_R$, which contributes to the potential of the
pseudo Goldstone bosons, 
has to be proportional to the breaking parameter $m^2$ and
thus can at most be logarithmically divergent. 

\begin{figure}
\begin{center}
\resizebox{5.in}{!}{\includegraphics*[100,450][450,750]{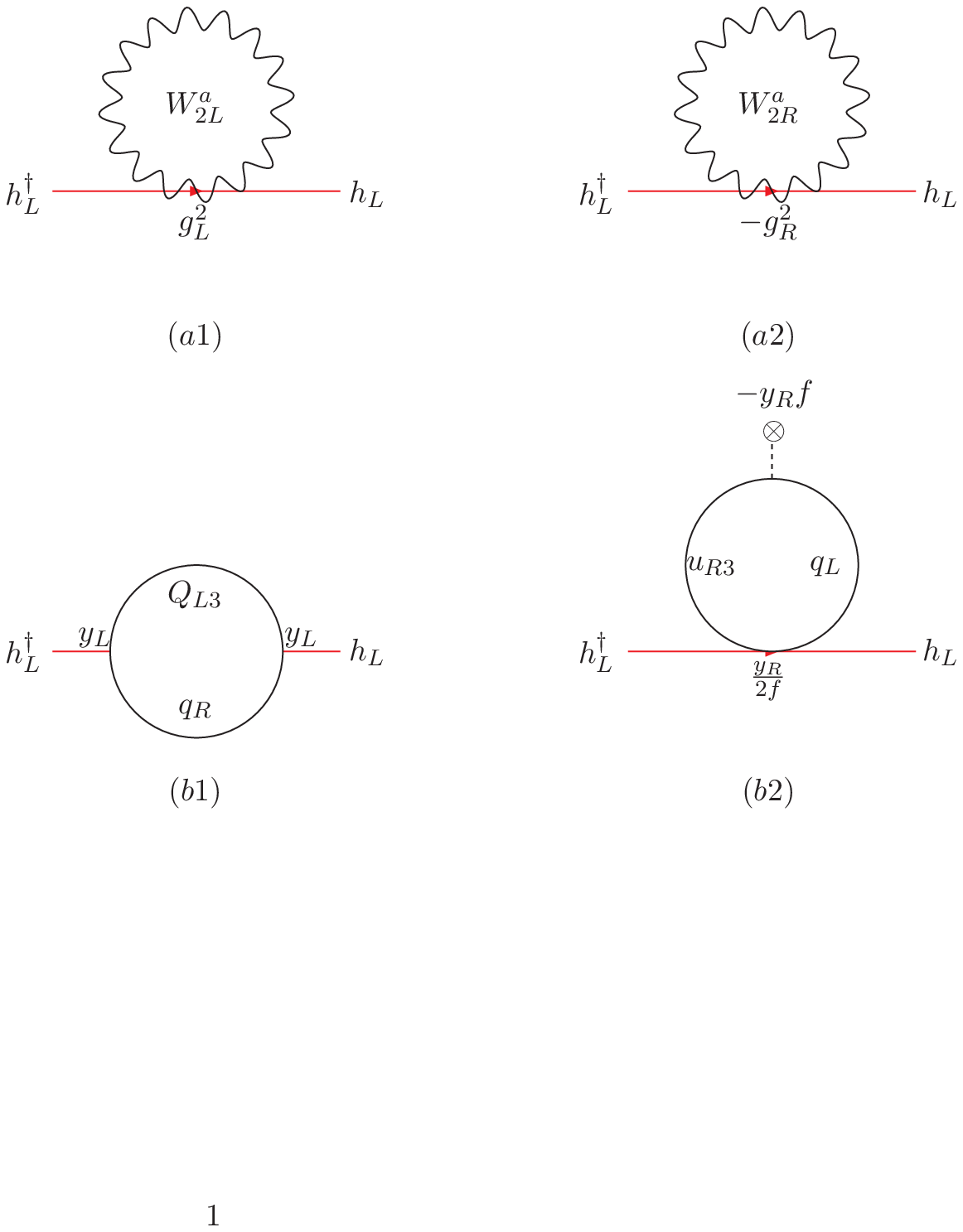}}
\caption{Diagrams responsible for the cancellation of one loop
quadratically divergent contributions to the Higgs mass. Diagrams 
(a1) and (a2) are contributions from ${\rm SU}(2)_L$ and ${\rm SU}(2)_R$
gauge bosons respectively. Diagrams (b1) and (b2) are contributions from light top and heavy
top respectively. }
\label{fig:cancellation}
\end{center}
\end{figure}

These cancellations can
be shown explicitly at one loop by using the Lagrangian 
and the expressions for the nonlinear Higgs fields given in the
previous sections. As we are to demonstrate the cancellation of
the quadratic divergence, we can treat all particles as
massless and ignore the mixing between particles.
The nonlinear Higgs fields $H_L$ and $H_R$ can be expanded as
\begin{equation}
H_L=h_L+\cdots,\ \ \ 
H_R=\left(
\begin{array}{c}
0\\
f-\frac{h_L^\dagger h_L}{2f}
\end{array}
\right)+\cdots\ .
\end{equation}
Let us first examine the contributions from the
${\rm SU}(2)_{L,R}$ gauge interactions.   The result can be easily
extended to the case of ${\rm U}(1)$.  Actually, the ${\rm U}(1)_{B-L}$
preserves the global U(4) symmetry and should not by itself
contribute to the Goldstone boson potential at all. 
For the gauge field loop contributions, 
the relevant vertices come from the gauge
interactions:
\begin{eqnarray}
    &&|D_{\mu} H_L|^2 + |D_{\mu} H_R|^2  \rightarrow
    g^2_{2L}H_L^{\dagger}(W_{2L}^{\dagger}W_{2L})H_L
+   g^2_{2R}H_R^{\dagger}(W_{2R}^{\dagger}W_{2R})H_R
\nonumber \\
&&=    g^2_{2L}h_L^{\dagger}(W_{2L}^{\dagger}W_{2L})h_L
+g^2_{2R}(0,f-\frac{h_L^\dagger h_L}{2f})
(W_{2R}^{\dagger}W_{2R})
\left(
\begin{array}{c}
0\\
f-\frac{h_L^\dagger h_L}{2f}
\end{array}
\right)+\cdots
\nonumber \\
&&=    g^2_{2L}h_L^{\dagger}(W_{2L}^{\dagger}W_{2L})h_L
-g^2_{2R}(h_L^{\dagger}h_L)(W_{2R}^{\dagger}W_{2R})_{22}+\cdots
\nonumber \\
&&=   \frac{1}{4} g^2_{2L}h_L^{\dagger}h_L
(W_{2L}^{a\dagger} W_{2L}^a)
-\frac{1}{4}g^2_{2R}(h_L^{\dagger}h_L)(W_{2R}^{a \dagger}W_{2R}^a)+\cdots
\end{eqnarray}
where the gauge fields 
$W_{2L,2R}=W_{2L,2R}^a\frac{\sigma^a}{2}$, and 
the subscript (22) denotes the (2,2) component of the
$W_{2R}^{\dagger}W_{2R}$.
The first term generates
a diagram as shown in figure \ref{fig:cancellation}.(a1) and the
second term generates a diagram as shown in figure
\ref{fig:cancellation}.(a2). 
If $g_{2L}=g_{2R}$, the two diagrams
give the same amplitude and cancel each other exactly due to the minus
sign of the second term. 

For the top loop contributions, the relevant vertices come from the Yukawa
interactions
\begin{eqnarray}
    &&y_L\bar{Q}_{L3}\tau_2 H_L^*q_R +y_R\bar{Q}_{R3}\tau_2H_R^*q_L+
    h.c.\nonumber\\
    &\sim & y_L\bar{Q}_{L3}\tau_2
    h_L^*q_R-y_R\bar{u}_{R3}q_L(f-\frac{h_L^{\dagger}h_L}{2f}+...)
+h.c.+\cdots
\nonumber \\
    &= & y_L\bar{Q}_{L3}\tau_2
    h_L^*q_R-y_R f\bar{u}_{R3}q_L + 
\frac{y_R}{2f}\bar{u}_{R3}q_L{h_L^{\dagger}h_L}
+h.c.+\cdots\ .
\end{eqnarray}
The first term
generates the usual diagram as shown in figure.
\ref{fig:cancellation}.(b1), with a contribution
proportional to $-y_L^2$.
The third term generates a diagram as shown in figure.
\ref{fig:cancellation}.(b2), with a insertion of
$-y_Rf\bar{u}_{R3}q_L$,  which is necessary as we have no
$\bar{u}_{R3}q_L$ propagator in the massless limit. Such diagram gives
a contribution proportional to
$-\frac{y_R}{2f}\times (-y_R f)\times 2$, where the factor of two takes into account the 
contribution from  the third term and its Hermitian conjugate.
The quadratic divergences in Fig.~\ref{fig:cancellation}.(b1) and (b2) 
cancel each other if $y_L=y_R$.

\section{Mass spectrum}
\label{sec:spectrum}
The new particles in the LRTH model are: 
heavy gauge bosons $Z_H$, $W_H^\pm$, heavy top quark $T$, 
neutral Higgs $\phi^0$, a pair of charged Higgses $\phi^\pm$,
and a ${\rm SU}(2)_L$ complex Higgs doublet: $\hat{h}_1^\pm$,
$\hat{h}_2^0$.  The model parameters are 
the Higgs vevs $f$, $\hat{f}$, the top quark Yukawa $y$, the cut off scale 
$\Lambda$, the top quark vector singlet mass mixing parameter $M$, a mass
parameter $\mu_r$ for $\phi^0$, and a 
mass parameter $\hat\mu$ for $ \hat{h}_1^\pm$ and
$\hat{h}_2^0$.  Once $f$ is fixed, the vev $\hat{f}$ can be 
determined by minimizing the CW potential 
for the SM Higgs and requiring that the SM Higgs obtains an electroweak
symmetry breaking vev of 246 GeV.  The top Yukawa $y$ can  
be fixed by the light top quark mass.  The remaining free parameters are:
($f$, $\Lambda$, $M$, $\mu_r$ and $\hat\mu$).

The value of $f$ and $\hat{f}$ are bounded from below by electroweak precision measurements,
which will be discussed in sec.~\ref{sec:constraints}.  It 
cannot be too large either since the fine tuning is more severe for 
larger
$f$.  In our analysis below, we pick $f$ to be in the range of 
500 GeV $-$ 1.5 TeV.  The corresponding fine tuning is in the range 
of 27\% to 4\%.  The cut off scale $\Lambda$ is typically chosen to be 
$4 \pi f$.  Sometime $\Lambda=2 \pi f$ is also considered.  
The mass mixing between the vector top single, $M$, controls the amount of 
${\rm SU}(2)_L$ singlet $q_L$ in the SM-like light top $t$.
It is therefore constrained   by the $Z\bar{b}b$ coupling 
and oblique parameters.
On the other hand, nothing forbids $M$ to 
be set to zero, which corresponds to zero mixing
between the light top quark and heavy top quark. 
The collider phenomenology for $M=0$ or very small value of $M$ ($\lesssim$
1 GeV)  differs dramatically from larger value of $M$, which will
be discussed separately in Sec.~\ref{sec:M0}.  The value for 
$\mu_r$ is non-zero, otherwise the neutral Higgs $\phi^0$ is 
massless.  The value of $\mu_r$ cannot be too large either, since otherwise the fine tuning of the SM Higgs
mass becomes severe.
In our analysis, we pick $\mu_r$ to be small, as the current experimental 
bound on the mass of $\phi^0$ is fairly weak.
The parameter $\hat\mu$ sets the masses for the Higgses $\hat{h}_1^\pm$,
$\hat{h}_2^0$.  Such a mass term breaks the left-right symmetry softly,
and could be of the order of $f$.  Although it is not particularly  relevant for collider
studies, it controls the mass of the dark matter candidate $\hat{h}_2^0$
and plays an important role in  the dark matter relic 
density analysis \cite{suDM}.

\begin{figure}
\begin{center}
%\resizebox{6.5 in}{!}{
\includegraphics*[width=3 in]{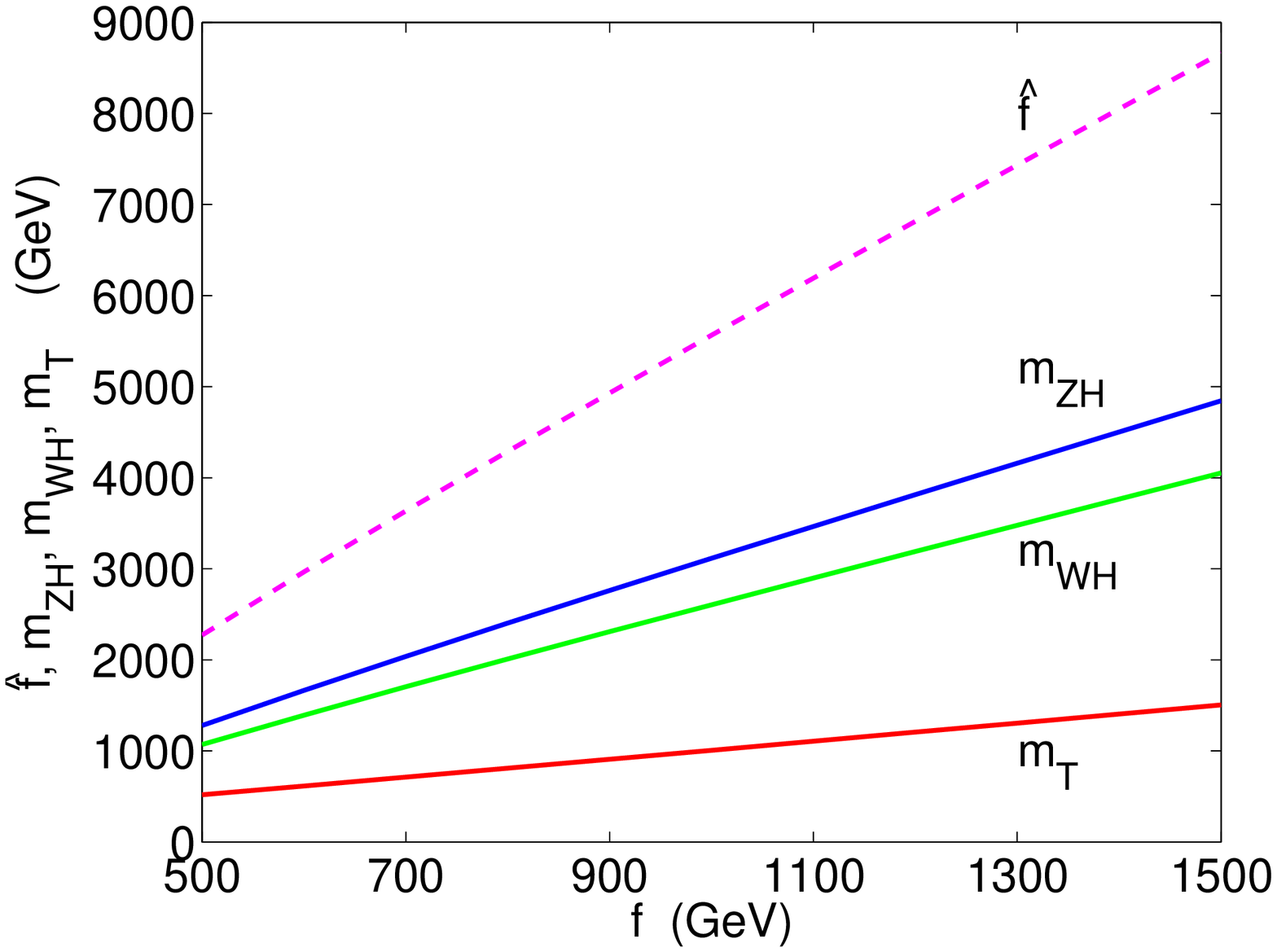}
\includegraphics*[width=3 in]{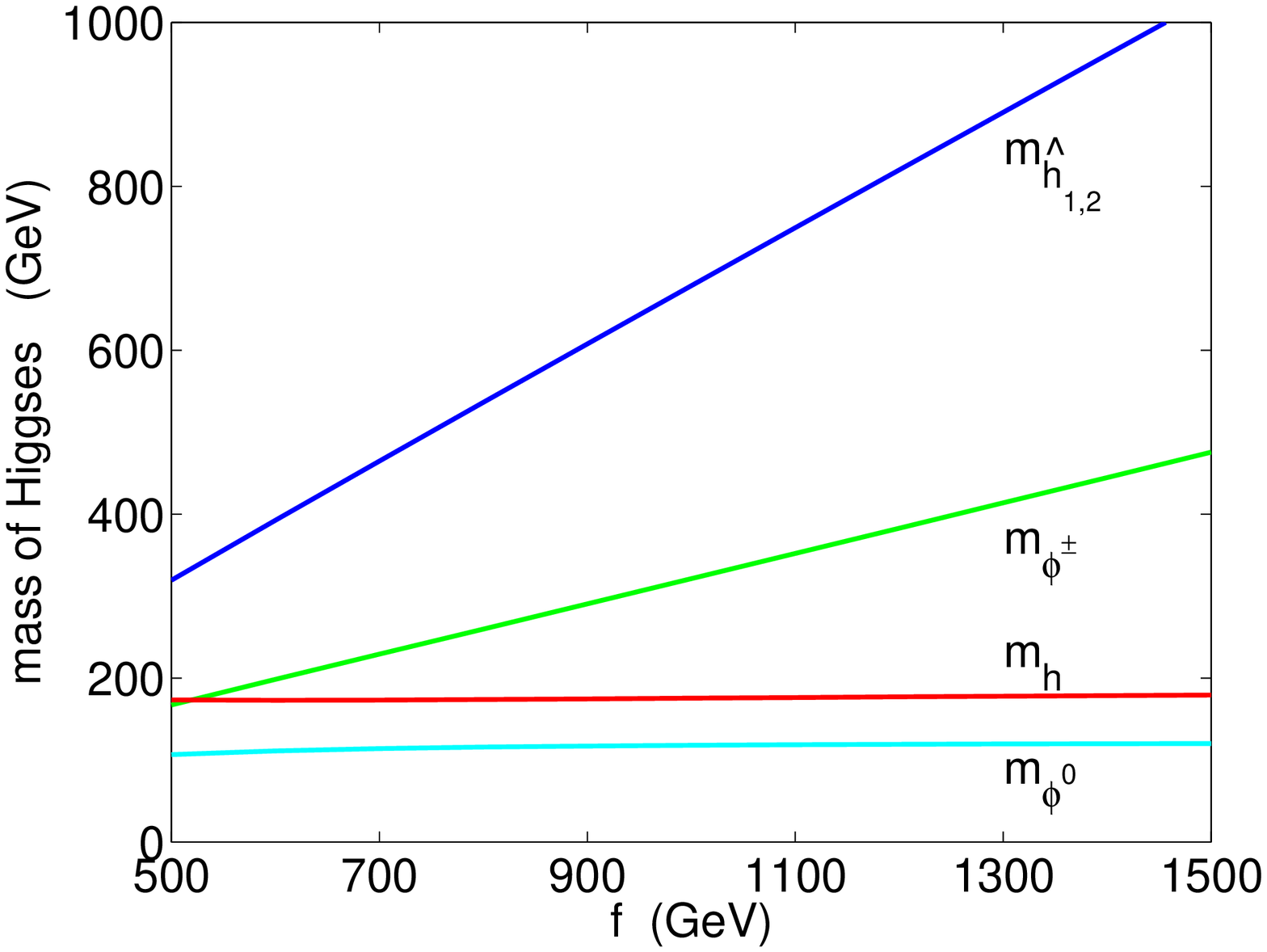}
%}
\end{center}
\caption{The left plot shows the 
value of $\hat{f}$ and masses of $Z_H$, $W_H$ and $T$.  The right plot 
shows the masses of $\hat{h}_1^{\pm}$ ($\hat{h}_2^{0}$), $\phi^\pm$,
$h$ and $\phi^0$.  The other parameters are chosen as
$\Lambda=4 \pi f$,  $M=150$ GeV, $\mu_r=50$ GeV and $\hat\mu=f/2$.
}
\label{fig:mass}
\end{figure}

Fig.~\ref{fig:mass} shows the masses of the new particles as
a function of $f$, for a typical set of parameter choices of
$\Lambda=4 \pi f$,
$M=150$ GeV, $\mu_r=50$ GeV and 
$\hat\mu=f/2$.  
The top curve in the left plot of Fig.~\ref{fig:mass}
shows the value of $\hat{f}$ as a function of $f$, which is determined from
the minimization of the CW potential of the SM Higgs.
The heavy top mass, which is determined by $f$, is 
between 500 GeV and 1.5 TeV.  The heavy gauge boson 
masses are above 1 TeV, heavier than the heavy top. 
This is because 
the heavy gauge boson masses are controlled by a much larger vev $\hat{f}$.
This mass hierarchy is different from 
the spectrum of the littlest Higgs model \cite{littest}, where the heavy top is 
heavier than the heavy $W_H$ \cite{lhtao}.  The masses of 
$W_H$ and $Z_H$ in the LRTH model are related: 
$m_{W_H}=m_{Z_H}\sqrt{\cos2 \theta_w}/\cos\theta_w$.  This mass relation is 
also different from the littlest Higgs model, where $m_{W_H}=m_{Z_H}$.
Choosing $\Lambda=2 \pi f$ instead of $4 \pi f$ 
leads to larger values of $\hat{f}$.
The masses for $W_H$ and $Z_H$ also become heavier, 
due to their $\hat{f}$ dependence.
The mass of the heavy top remains unchanged, since
it is independent of $\hat{f}$. 
All of those particles are within the reach of the LHC.

The right plot of Fig.~\ref{fig:mass} shows the masses for all the Higgses 
in the LRTH model.
The mass of the Higgs $\phi^0$ is related to $\mu_r$ as
$m_{\phi^0}=\mu_r\sqrt{\hat{f}/f}$.  For $\mu_r=50$ GeV, the mass for 
$\phi^0$ is around 100 GeV.  The masses of the charged Higgses $\phi^\pm$ obtain 
contributions from both the $\mu_r$ term, similar to
the neutral Higgs $\phi^0$, and the CW potential, 
$(g^4/16 \pi^2) \hat{f}^2 
\ln \Lambda^2/(g^2\hat{f}^2)$. 
Their  masses increase with $f$, and 
are between 200 to 400 GeV.   
The SM Higgs mass is determined 
by the CW potential.  It varies between 145 $-$ 180 GeV, 
depending slightly on the values of $\Lambda$  and $M$.  
The masses of the Higgses $\hat{h}_1^\pm$ and 
$\hat{h}_2^0$ are nearly degenerate, with a small splitting caused by the 
electromagnetic interactions.  Three individual pieces contribute to 
its mass squared: $\hat{\mu}^2$, $\mu_r^2(f/\hat{f})$ 
and terms from the CW potential.
The CW contribution is between $(200 {\rm \ GeV})^2$ to 
$(700 {\rm \ GeV})^2$  for $f$ varies between 500 GeV to 1500 GeV.
For smaller values of $\Lambda$, all the Higgs masses except 
$\phi^0$ decrease. 
For $\phi^0$, the mass increases slightly, due to the 
larger value of $\hat{f}$.
the LHC reach of these particles depends on 
their production processes and decay modes, which will be discussed 
in Sec.~\ref{sec:signal}.

For smaller value of $M$, $\hat{f}$ decreases.  This leads to 
a slightly smaller value for $m_{W_H}$, $m_{Z_H}$ and all the Higgs masses. 
The heavy top mass also decreases  due to the smaller splitting between the 
light and heavy tops. 

\section{Experimental constraints}
\label{sec:constraints}

The strongest experimental constraints on the LRTH model come from the precision 
measurements on the virtual effects of heavy gauge bosons, and 
the mass bounds from the direct searches at high energy colliders.

The constraints on the mass of the heavy $W_H$ depend on the masses of the 
right handed neutrinos.  For $m_{\nu_R}<m_e$, $m_{W_H}$ is constrained to be 
larger than 4 TeV to avoid the over production of ${}^4{\rm He}$ \cite{nucleo}.
For $m_{\nu_{eR}}<m_p$, supernova cooling constrains $m_{W_H}$ 
to be larger than 23 TeV \cite{supernova}.  However, once the right handed 
neutrinos are heavy, all those constraints are relaxed.  In the LRTH, the 
right handed neutrinos could obtain large Majorana masses of the order of 
$\hat{f}^2/\Lambda$, and 
the above mentioned constraints on $m_{W_H}$ are therefore absent.  The strongest
constraint on $m_{W_H}$ then comes from $K_L-K_S$ mixing.  The box diagram
with the exchange of one $W$ and one $W_H$ has an anomalous enhancement 
and yields the bound $m_{W_H}>1.6$ TeV \cite{KLKS}, 
which translates into a lower limit 
on $f$ to be 670 GeV.  This analysis, however, did not include
higher order QCD corrections and it used vacuum insertion to obtain the matrix
element.  An update on the 
$m_{W_H}$ constraints from $K_L-K_S$ mixing is under current 
investigation \cite{KLKSupdate}.  The current limit also assumes that the CKM 
matrix for the right handed quark sector is the same as or the complex 
conjugate of the one for the left handed quark sector.  The bound
on $m_{W_H}$ can further be relaxed if we drop this assumption.   It will
lead to a breaking  of left-right symmetry in the first two generations.
This is safe since  no large contributions  to the Higgs masses 
appear from the first two generation quarks 
due to the smallness of their Yukawa couplings. 
The direct search limit on $m_{W_H}$ depends on the masses of $\nu_R$. 
If  $m_{\nu_R}>m_{W_H}$, $W_H\rightarrow 
l \nu_R$ is forbidden.  D0 excludes the mass range of 300 to 800 GeV
assuming $W_H$ decays dominantly into two jets \cite{D0WH}.  
The CDF excludes the mass region of 225 to 
566 GeV in $t\bar{b}$ final states~\cite{CDFWH}.  The CDF bound is weaker
for the heavy $W_H$ in the LRTH since the decay of $W_H\rightarrow {t}\bar{b}$
is suppressed by the smallness of $M$.
For $m_{\nu_R}\ll m_{W_H}$, CDF finds $m_{W_H}>786$ GeV using the $e$ and $\mu$ 
final states combined \cite{CDFnuR}, while the D0 limit is 720 GeV \cite{D0nuR}.
For $m_{\nu_R}= m_{W_H}/2$, the D0 bound weakens to 650 GeV~\cite{D0nuR}. 

Unlike the heavy charged gauge boson $W_H$, which does not mix with the 
SM $W$, the heavy $Z_H$ mixes with the SM $Z$ with a mixing angle of the 
order of $v^2/\hat{f}^2$. There are three types of indirect constraints.  
$Z$-pole precision measurements constrain only the $Z-Z_H$ mixing.  The 
low energy neutral current processes and high energy precision measurements
off the $Z$-pole are sensitive not only to $Z-Z_H$ mixing, but also to the
direct $Z_H$ exchange.  The limit on the $Z-Z_H$ mixing is typically $<$
few $\times 10^{-3}$ \cite{PDG}, 
translating into $\hat{f}$ ($f$) to be larger than a few TeV (500$-$600 GeV). 
The lower bound on the heavy $Z_H$ mass from precision measurements 
is about 500$-$800 GeV \cite{PDG}.
$Z_H$ can also be directly produced at high energy colliders and decays
into quarks or leptons.  In the leptonic final states, the current bounds from CDF is about 
630 GeV \cite{PDG}.  

\section{Sketches for Future Collider  phenomenology}
\label{sec:signal}

In this section, we discuss the collider phenomenology of the new
particles in the LRTH model.  We present
the production cross sections  and particle decay branching ratios.  
All the numerical studies are done using 
CalcHEP \cite{calchep}. 
Signals typically involve multijets, energetic leptons and missing energies.
The SM backgrounds are in general unsuppressed, and more detailed analyses 
for individual processes are needed to identify the discovery potential for the 
LRTH model at the LHC.  Such study is beyond the scope of the current paper and 
we leave it for future work \cite{collidersu}.
%Since the signals generally involve more than one 
%$b$-jet, the use of multiple $b$-tagging can provide rejection of the 
%jet background.  In addition, the reconstruction of the intermediate on-shell
%states is a useful tool to distinguish  the signals against the background.
%More detailed studies of the discovery potential for LRTH at the LHC including background %analyses are left for future work\cite{collidersu}.

Since the decays of the particles depend on the left-right mass mixing of 
the top singlet $M \bar{q}_L q_R$, which therefore changes the collider signals, 
we first discuss the general case with a small $M$, choosing 
$M=150$ GeV as an illustration. For very small value of $M$ ($\lesssim$ 1 GeV),
in particular,  for $M=0$, the decay patterns of certain
particles change dramatically, which leads to completely different
collider signals.  We devote Sec.~\ref{sec:M0}
for the discussion of such case.

\subsection{Heavy top quark}
\label{sec:heavytop}

\begin{figure}
%\resizebox{7.5in}{!}
{\includegraphics*[0,600][450,720]{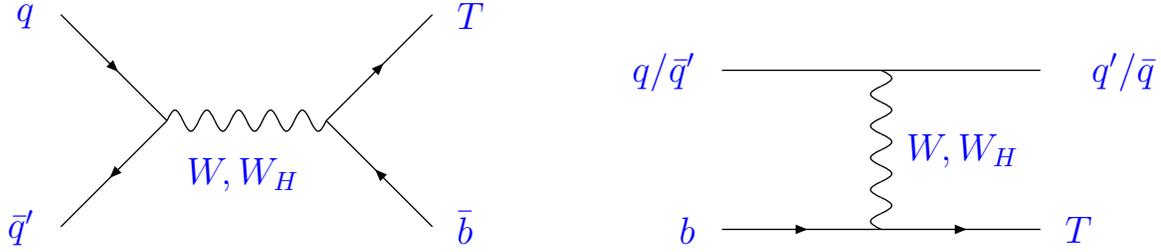}}
\caption{Feynman diagrams for single heavy top production.}
\label{fig:singletop}
\end{figure}

\begin{figure}[ht]
\resizebox{6.3 in}{!}{
\includegraphics{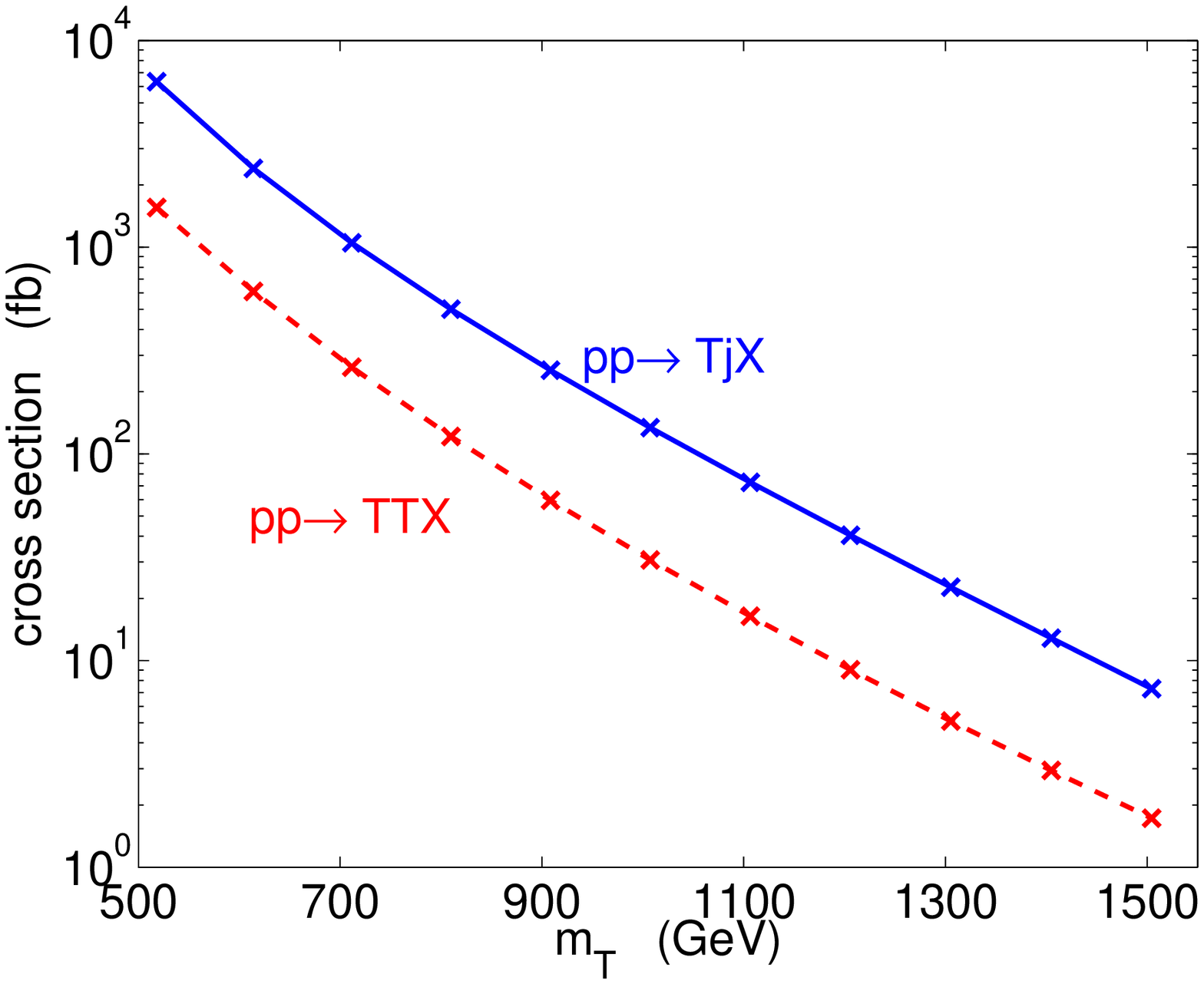}
\includegraphics{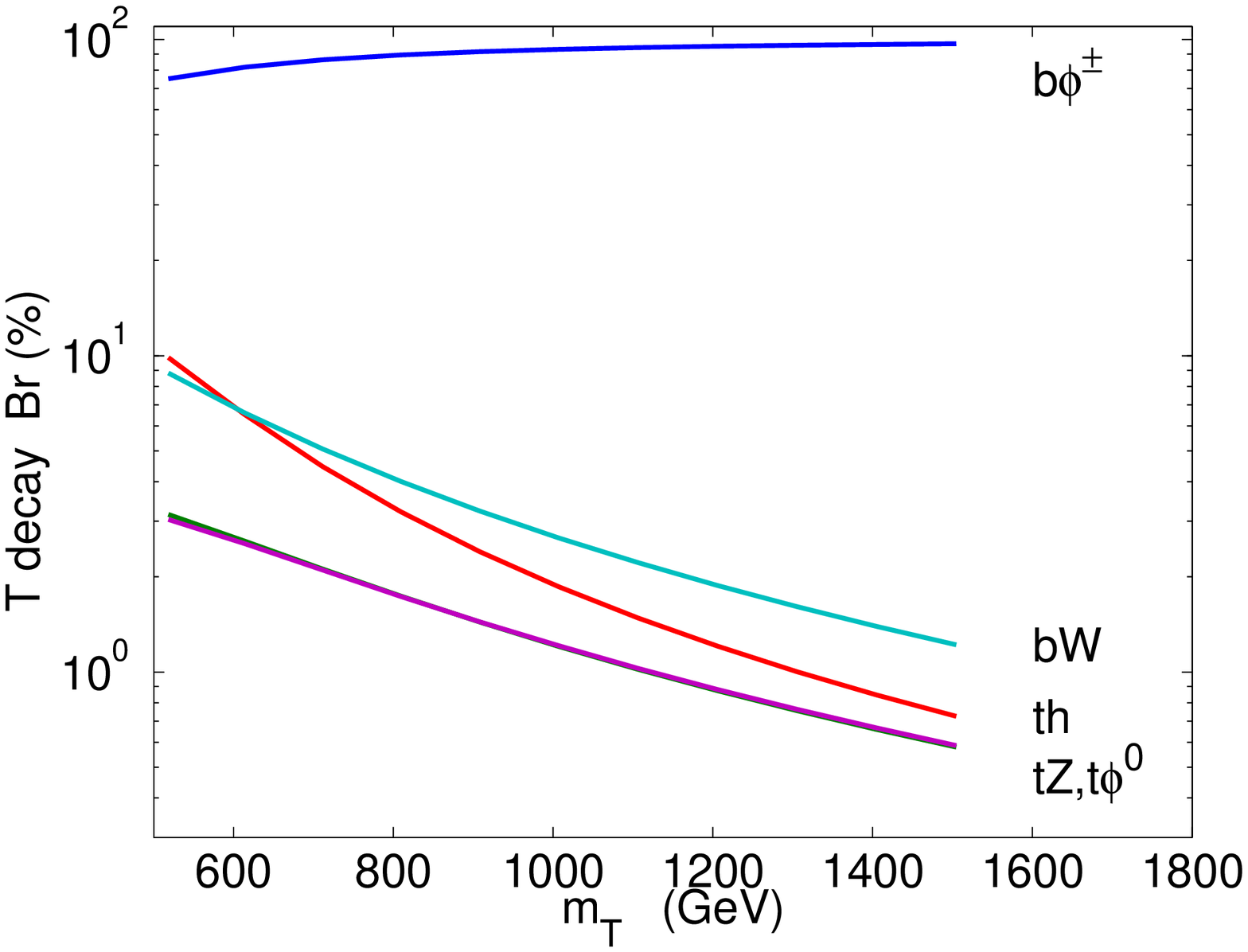}}
\caption{Left plot shows the single and pair production for heavy top quark
at the LHC.  The ``x'''s correspond to the 
value of $f$ being $500,600,...,1500$ GeV.
The right plot shows the branching ratios of the heavy top decay for
$M=150$ GeV.
}
\label{fig:heavytop}
\end{figure}

A single heavy top quark can be produced at the LHC dominantly via $s$-channel or $t$-channel
$W$ or $W_H$ exchange, as shown in Fig.~\ref{fig:singletop}.  The associated 
jet is mostly a $b$-jet in the former case, or $u/d$ jets in the latter
case.  Since $W_H$ is heavier than $T$ in the LRTH models,
the $s$-channel on-shell $W_H$ decay dominates the single heavy top production,
contributing to more than 80\% of the total cross section.
The contribution from $W$ boson exchange is negligible, since $W\bar{T}b$ 
coupling is suppressed by $(M/f)(v/f)$, which vanishes in the limit of $M=0$.
This is different from the little Higgs model, where the $t$-channel
$W$ exchange dominates the single heavy top production cross section.
% both LH and TH has v/f suppression.  TH has additional M/f suppression.

The single heavy top quark production 
cross section is shown by the solid curve 
in the left plot of Fig.~\ref{fig:heavytop}.  
For a heavy top mass of 500$-$1500 GeV, the cross section is in the 
range of $7\times 10^3$ fb $-$ 10 fb. 
It is comparable to the single 
heavy top production cross section 
in the littlest Higgs model \cite{lhtao}, which is 
about 20 fb for a 1500 GeV heavy top.   
We also show  the cross section of heavy top 
pair production (dashed line in the left plot of Fig.~\ref{fig:heavytop}).
The dominant contribution comes from gluon exchange:
$q\bar{q}, gg \rightarrow T \bar{T}$.  Although the QCD coupling is larger,
this channel suffers from the phase space suppression due to the large 
heavy top mass.  The cross section 
is about a factor of five smaller when compared to the single 
heavy top production mode.

The decay branching ratios of the heavy top are shown in the
right plot of Fig.~\ref{fig:heavytop}.
For $M=150$ GeV, more than 70\% of heavy top decays via 
\beq
T \rightarrow \phi^+ + b,
\eeq
with a partial decay width of 
\beq
\Gamma(T\rightarrow \phi^+ b)=\frac{1}{8 \pi}\frac{p_b}{m_T}
\left[
{E_b}(|g_L|^2+|g_R|^2)+{m_b}(g_Lg_R^*+g_L^*g_R)
\right],
\eeq
where $p_b$  and $E_b$ is the momentum and energy 
of $b$-jet in the rest frame of the heavy 
top and $g_L$ and $g_R$ are the left and right handed couplings 
of $\phi^-\bar{b}T$:   $\phi^-\bar{b}(g_Lp_L+g_Rp_R)T$, 
which can be read off from Table~\ref{tab:higgsfermion}. 
In the limit of $M\ll f$,
$g_L \sim i y $ and $g_R \sim im_b/f \sim 0$, the partial decay
width simplifies to 
\beq
\Gamma(T\rightarrow \phi^+ b)=\frac{y^2}{16 \pi}p_b\approx 
\frac{y^2}{32 \pi}m_{T}.
\label{eq:tHdecay}
\eeq
In the last step, we have ignored the final state masses since they are small 
compared  to large $m_T$.

Considering the subsequent decay of 
\beq
\phi^+ \rightarrow tb, \ \ 
t \rightarrow W^+ b \rightarrow l^+ \nu b,
\label{eq:tHtobphi}
\eeq
the signal is 3 $b$-jets $+$ one charged lepton ($e$ or $\mu$) 
$+$ missing $E_T$.  
There is always an additional  energetic 
jet (most likely a $b$-jet) that accompanies $T$ from single heavy top production process.
Due to the large single heavy top production cross section and 
${\rm Br}(W\rightarrow e\nu_e, \mu\nu_{\mu})\sim 20\%$, 
more than 10,000 events can be seen 
with $10\ {\rm fb}^{-1}$ luminosity
for a heavy top mass of around 600 GeV.
The SM backgrounds come from $t\bar{t}$, $W+$4 jets and $tbj$. 
Preliminary study in Ref.~\cite{collidersu} shows that the jet associated with 
the single $T$ production is typically very energetic comparing to the jets
from $t\bar{t}$ decays.  A cut on the transverse momentum of the most  
energetic jet offers an effective way to suppressed the dominant 
$t\bar{t}$ background while retains most of the signals. 
In addition,   the reconstruction of $W$, $t$,
$\phi^+$ and $T$ 
can be used to discriminate the signal from the background.
We can reconstruct the
$W$ boson using the invariant mass of the lepton and neutrino 
\footnote{If the missing energy is solely due to the neutrino,
the neutrino momentum can be reconstructed with a two-fold ambiguity under the 
approximation that $m_\nu=0$.}.
Combining the $W$ with one $b$-jet, we require the invariant mass to be 
around the top quark mass.  Similarly, we can reconstruct $\phi^+$ through
the combination of $tb$ and reconstruct ${T}$ using $b\phi^+$.  

%Since we do not know the mass for $\phi^\pm$ and $T$ a priori, 
%and there are multiple $b$-jets in the final states, we might 
%suffer from the combinatorial problem of selecting the correct 
%combination of $b$-jets.
 
The heavy top can also decay into $ht$, $Zt$ and 
$Wb$.  The decay branching ratios are suppressed since the relevant 
couplings are suppressed by at least one power of $M/f$.
The $\bar{T}_Rt_{L}h$ and $\bar{T}_Rt_{R}Z$ couplings are 
proportional to the fraction
of $q_R$ in $T_R$, which is about $M/f$.
The $\bar{T}_Lt_{R}h$, $\bar{T}_Lt_{L}Z$ 
and $\bar{T}_Lb_LW$ couplings are 
proportional to the fraction of $u_{L3}$ in $T_L$, which is about $
(M/f)(v/f)$.  
For large $m_{T}$, the relation 
\beq
\Gamma(T\rightarrow ht)=
\Gamma(T\rightarrow Zt)=
\frac{1}{2}\Gamma(T\rightarrow Wb)
\eeq
still holds as in the littlest Higgs models \cite{lhtao}, 
due to the Goldstone boson equivalence theorem.  
However, such relation is hard to test at the LHC because of the 
suppressed branching ratios into those channels.
For $M=150$ GeV, 
the branching ratio for $T\rightarrow Wb$ is about 10\% for $m_T\sim$ 
500 GeV, and decreases quickly for larger $m_T$.

The search for $T\rightarrow Wb$ 
is similar to the usual single top quark searches \cite{atlasTDR, singletop}. 
The leptonic $W$ decay yields a nice signal of 
one $b$-jet $+$ one electron or muon $+$ missing $E_T$. 
For single $T$ production channel, there is usually additional energetic jet which is most 
likely a $b$-jet. 
Requiring one energetic lepton, at least two energetic jets and at 
least one energetic $b$-tagging jet reduces the enormous QCD multijet
background \cite{atlasTDR}.   The remaining dominant SM backgrounds are 
SM single top production(via $Wt$, $W$-gluon fusion or $W^*$ processes),   
$t\bar{t}$, $Wb\bar{b}$ and $Wjj$.  
Studies in Ref.~\cite{singletop} shows that requiring no more than two jets
can be used to reduce the $t\bar{t}$ and $Wt$ background,
which has on average more jets than the single heavy top process. 
Requiring more than one  $b$-tagging jet reduces the $Wjj$ and 
the SM $Wt$ and $W-$gluon fusion background.  
Since the neutrino momentum can be fully reconstructed (with a two-fold ambiguity), requiring 
$m_{l\nu b}$ to lie around $m_T$ reduces $Wjj$, $Wbb$ and single top background. 
Further rejection of $Wjj$ and $Wbb$ background can be achieved by impose a cut on the 
scalar sum of the jet $p_T$, which typically has a lower value.
Similar analysis for $T\rightarrow Wb$ in the little Higgs models has been studied in 
Ref.~\cite{LHCTH}, and it was shown that for ${\cal L}=300\ {\rm fb}^{-1}$, 5 $\sigma$ discovery is possible for $m_T$ up to about 2 TeV.   Note,  however,  that in the little Higgs model, ${\rm Br}(T\rightarrow Wb)=50$\%, while in LRTH, the branching ratio is much less, 
depending on the values of $M$ and $f$.

At small value  of $f$ around 500 GeV, the branching ratio for $T\rightarrow ht$ is about 
10\%.  Since the mass of $h$ in the LRTH models is typically around 170 GeV, it decays dominantly into $WW^*$ or $ZZ^*$, leading to multilepton signals.    The main background is 
top pair production, where both tops decay semileptonically and a third lepton can arise from a $b$ jet.   Studies for channels with similar final states in the little Higgs models (for $V_H\rightarrow Vh$ with a heavy $h$) have been discussed in  Ref.~\cite{LHCLHVH}.

The heavy top can also decay into $Zt$:  
\beq
T\rightarrow Z + t,\ {\rm with\ } Z \rightarrow l^+ l^-, \ 
{\rm and\ } t \rightarrow W^+ b \rightarrow l^+ \nu b.
\label{eq:tHtotZ}
\eeq
The signal is 1 $b$-jet $+$ tri-lepton $+$ missing $E_T$.  
The dominant SM background comes from $WZ$, $ZZ$ and $tbZ$.  
Similar studies in the framework of little Higgs models \cite{LHCTH} show
that requiring three isolated energetic lepton (either $e $ or $\mu$), energetic $b$-jets, missing $E_T$ larger than 100 GeV, and a pair of leptons with reconstructed invariant mass
around $m_Z$ rejects most of the background.  At ${\cal L}=300\ {\rm fb}^{-1}$,  5 $\sigma$
discovery at the LHC is possible for $m_T$ up to about 1 TeV 
(with ${\rm Br}(T\rightarrow Zt)=25$\%).  In the LRTH, however, such channel is only useful 
for small $f$ and not so small $M$. 

The decay of 
\beq
T\rightarrow t + \phi^0,\ {\rm with\ } \phi^0 \rightarrow b\bar{b}, \ 
{\rm and\ } t \rightarrow W^+ b \rightarrow l^+ \nu b.
\label{eq:tHtotphi}
\eeq
is also possible for small values of $f$.  The signal is three $b$-jets, plus energetic lepton and missing $E_T$.  Such process is very similar to 
$T\rightarrow ht$ with $h\rightarrow b\bar{b}$ in the little Higgs models \cite{LHCTH}.
The dominating background comes from $t\bar{t}$, which can only be distinguished by studying the kinematics.  Studies \cite{LHCTH} showed that the discovery in such mode is more difficult comparing to $Wb$ and $Zt$ modes that we discussed before.  It can, however, be used as an confirmation if heavy top partners are discovered in other channels.

Due to the small 
mixing of the vector top singlet, the deviation of $Wtb$ coupling
from its SM value is of the order of $(M/f)^2(v/f)^2$, which is usually
less than a few percent.  Such a small deviation is very hard to observe,
even at a high luminosity linear collider.

\subsection{Heavy gauge bosons}
\label{sec:heavygauge}

\begin{figure}
\resizebox{4 in}{!}{
\includegraphics{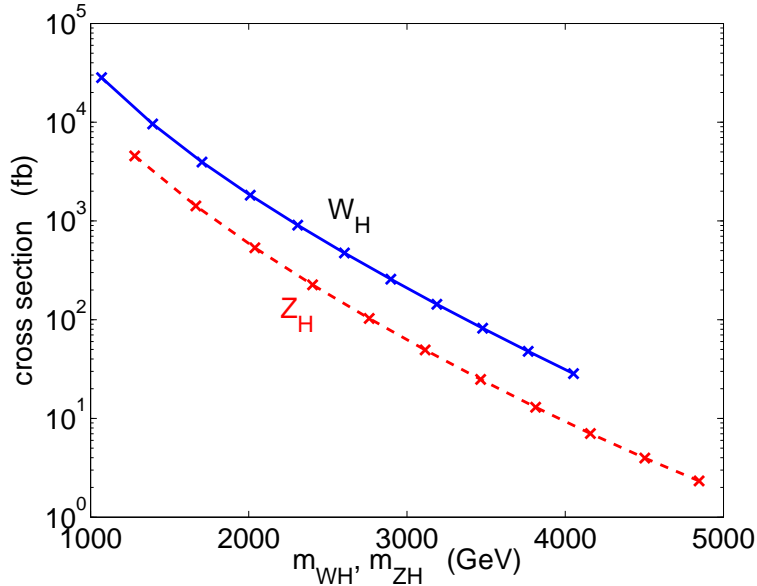}}
\caption{Cross section for heavy gauge bosons $W_H$ and $Z_H$ Drell-Yan 
productions at the LHC.
The ``x'''s correspond to the 
value of $f$ being $500,600,...,1500$ GeV.}
\label{fig:drellyan_WHZH}
\end{figure}

The dominant production channels for heavy gauge bosons at hadron colliders are
the Drell-Yan processes: $pp\rightarrow W_H X$ and $pp\rightarrow Z_H X$.
The production cross sections are shown in Fig.~\ref{fig:drellyan_WHZH}.
%Here we have used the narrow width approximation to  
%extract the Drell-Yan cross sections:
%\bea
%\sigma(pp\rightarrow W_H^+)&=&
%\frac{\sigma(pp\rightarrow W_H^+ \rightarrow u \bar{d})}
%{{\rm Br}(W_H^+\rightarrow u \bar{d})},\\ 
%\sigma(pp\rightarrow Z_H)&=&
%\frac{\sigma(pp\rightarrow Z_H \rightarrow e^+ e^-)}
%{{\rm Br}(W_H^+\rightarrow e^+ e^-)}.
%\eea
%Since $\Gamma_{W_H, Z_H} \ll m_{W_H, Z_H}$
%in left-right twin higgs model,
%this approximation only introduces a small error of the order of 
%$\Gamma/m$, which is about a few percent. 
%A QCD K-factor of $1+8\pi /9 \alpha_s$ has 
%been included in the numerical calculations.
The heavy right handed 
$W_H$ boson couples to the SM light quark pairs with the SM coupling
strength. The Drell-Yan cross section is large: 
varying from $3 \times 10^4$ fb 
for $W_H$ mass of about 1 TeV to 30 fb for $W_H$ mass of about 4 TeV.
For the heavy $Z_H$, the cross section is smaller comparing to $W_H$, 
due to the smaller $Z_H$ coupling to the SM fermion pairs as shown in 
Table~\ref{tab:gaugefermion}.  The cross section is still sizable: varying 
from $5\times 10^3$ fb for $Z_H$ mass around 1.3 TeV to 2 fb for $Z_H$ 
mass around 5 TeV. 

\begin{figure}
\resizebox{6.3 in}{!}{
\includegraphics{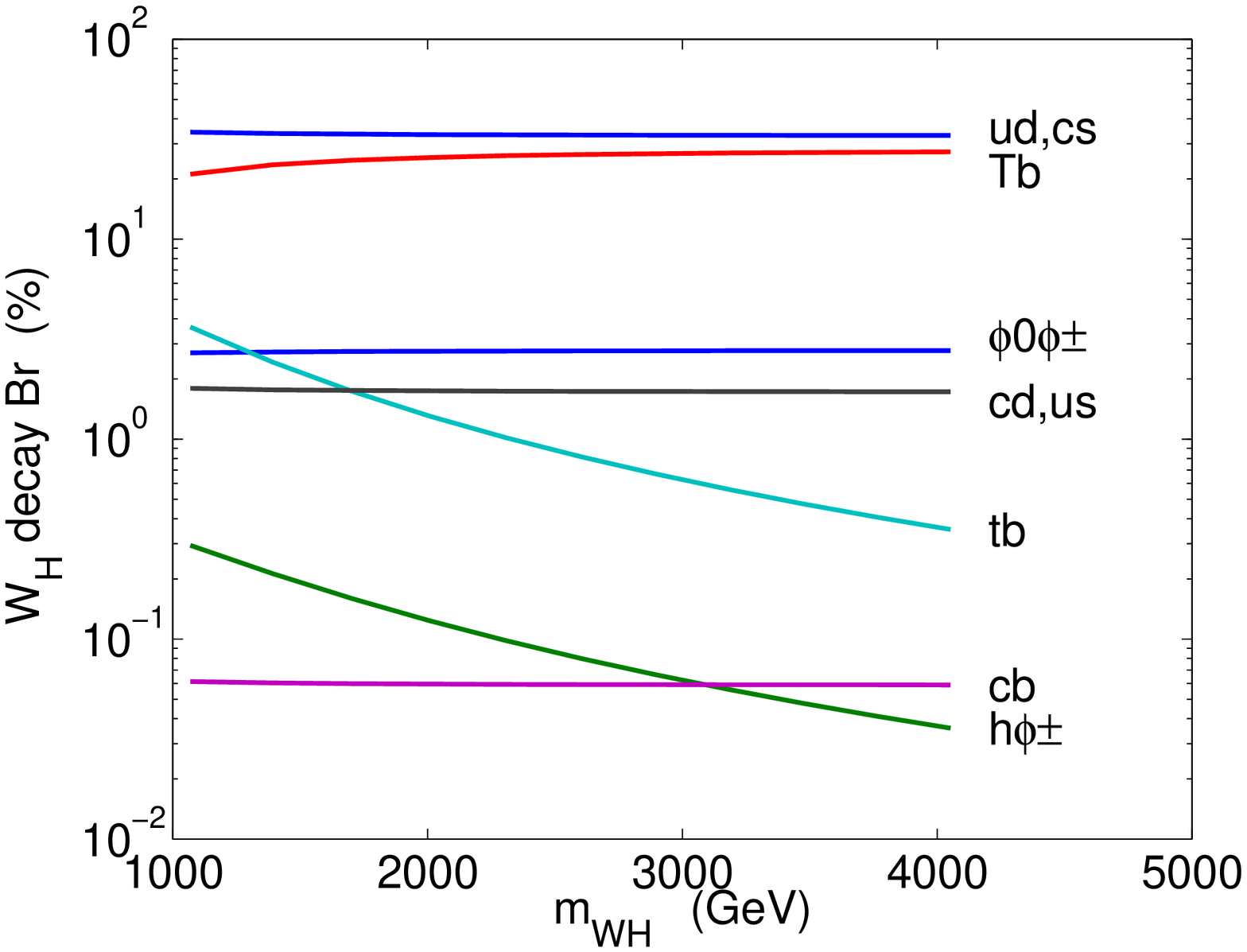}
\includegraphics{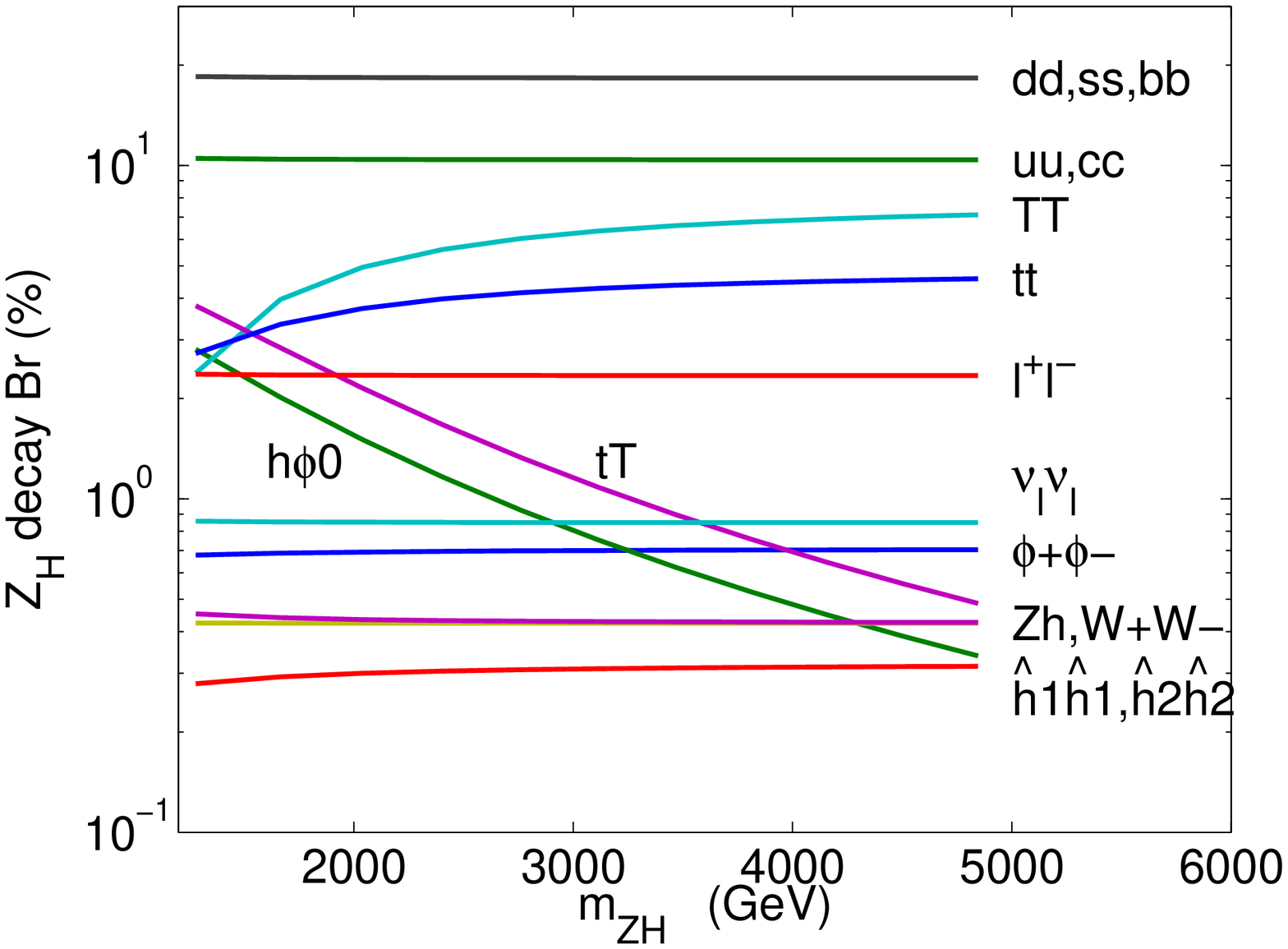}
}
\caption{Decay branching ratios of the heavy gauge bosons $W_H$ and $Z_H$ for
$M=150$ GeV.  
Here we have assumed that $m_{\nu_R}>m_{W_H}$ and the leptonic decay modes for $W_H$ are absent. }
\label{fig:decay_WHZH}
\end{figure}

In Fig.~\ref{fig:decay_WHZH}, we show the decay branching ratios of $W_H$ and
$Z_H$ as a function of the gauge boson masses.  For $W_H$, it can not decay to the SM leptons and neutrinos since it is a purely ${\rm SU}(2)_R$ gauge boson.  It could, however,
decay into $l+\nu_R$ if $m_{W_H}>m_{\nu_R}$,   which  will be discussed later. In Fig.~\ref{fig:decay_WHZH}, 
such leptonic decay 
mode is absent since the right handed neutrino masses are set to be larger than
$m_{W_H}$ in our analyses.
The dominant decay mode for $W_H$ is into two jets, 
with a branching ratio of about 30\%.  
Such mode suffers from the overwhelming QCD di-jets background
for large $p_T$ jets \cite{dijetLHC}.  
Current limits on dijets events from resonance 
decay \cite{RunIINP} is relatively weak.

$W_H$ could also decay into a heavy top plus a $b$-jet,
with a branching ratio of about 20\%$-$30\%.  Depending on the 
subsequent decays of the heavy top, we expect to see signals of
\begin{itemize}
\item{4 $b$ + lepton ($e$ or $\mu$) + missing $E_T$, with a 
branching ratio suppression factor of 
${\rm Br}(T \rightarrow \phi^+ b)\times{\rm Br}(W \rightarrow l^+ \nu)>14\%$.
The dominant SM backgrounds are $t\bar{t}$ and $W+jjjj$.}
\item{2 $b$ + lepton ($e$ or $\mu$) + missing $E_T$, with a 
branching ratio suppression factor of 
${\rm Br}(T \rightarrow W^+b)\times{\rm Br}(W \rightarrow l^+ \nu)<2\%$.
The dominant SM backgrounds are $tj$, $t\bar{t}$, $Wbb$ and $Wjj$.}
\item{2 $b$ + tri-lepton($e$ or $\mu$) + 
missing $E_T$, with a 
branching ratio suppression factor of 
${\rm Br}(T \rightarrow Z  t)\times{\rm Br}(W \rightarrow l^+  \nu)
\times{\rm Br}(Z \rightarrow l^+  l^-)<6\times 10^{-4}$. 
The dominant SM background is $tbZ$.}
%Although the signal is smaller due to the small
%leptonic $Z$, $W$ decay branching ratios and small 
%${\rm Br}(T \rightarrow Z t)$, the tri-lepton signal provides a very nice 
%suppression of the SM background.}
\end{itemize}
%The reconstruction of the heavy top and other 
%intermediate on-shell states, as explained in Sec.~\ref{sec:heavytop},
%would distinguish the signal from the background.
Since single $T$ production mostly comes from on-shell $W_H$ decay, the discussion in Sec.~\ref{sec:heavytop} for heavy top partners also applies to $W_H$ study here.

$W_H$ could also decay into $\phi^0\phi^\pm$ with a branching ratio of about 
3\%.  This is the dominant production mode for $\phi^0$.

The $W_H\rightarrow tb$ branching ratio is of the order of 4\% or less. Search of 
$tb$ final states from a heavy $W_H$ decay has been studied in Ref.~\cite{sullivan}.
It has been shown that at the LHC, with 10$-$ 100 ${\rm fb}^{-1}$ luminosity,
a reach of $m_{W_H}$ of $3-4$ TeV is possible at 95\% C.L.

For $1\ {\rm GeV}<m_{\nu_R}<m_{W_H}$, where the lower bound is imposed to 
avoid the strong constraints on the $W_H$ mass
from either supernova cooling \cite{supernova} or the relic 
abundance of ${}^4{\rm He}$,
$W_H\rightarrow l \nu_R$ is possible, with a branching ratio of about
9\%.   $\nu_R$ further decays into lepton plus jets.   The details of the decay process are 
very model dependent, which will not be further discussed here. 
%Dominating backgrounds are $W$+jets and $t\bar{t}$ events.  Studies for similar 
%final states have been performed in Ref.~\cite{WhZh}, ...
%The  
%reconstruction of  $\nu_R$ can be used to reconstruct the $W_H$ mass and distinguish the 
%signal from the background.  
%SAY SOMETHING HERE.

\begin{figure}
\resizebox{4 in}{!}{
\includegraphics{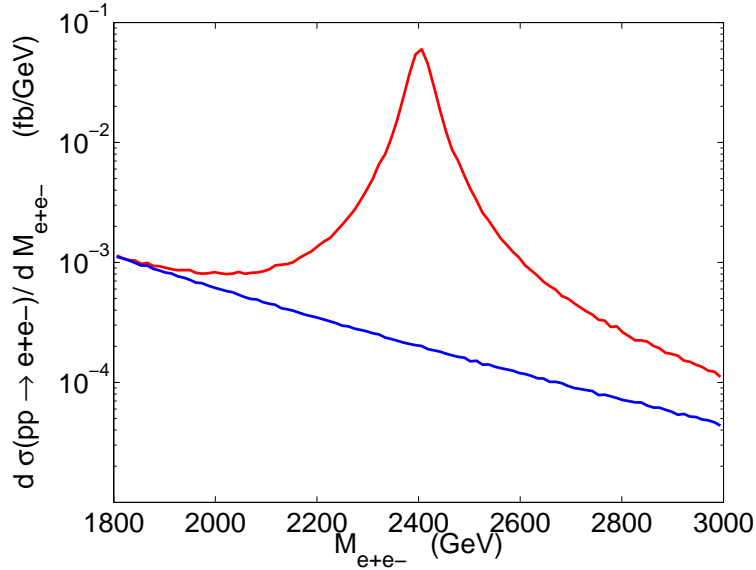}
}
\caption{The invariant $e^+e^-$ mass distribution at the LHC.  The
blue (dark) line is for the SM background, and the red(light) line
is for the LRTH model 
where a heavy $Z_H$ is produced through the
Drell-Yan process.  The model parameters are chosen to be 
$f=800$ GeV, $M=150$ GeV and $\Lambda=4 \pi f$.  The corresponding
$Z_H$ mass is 2403 GeV, with a decay width of $\Gamma_{Z_H}=51$ GeV.
 }
\label{fig:ZHeE}
\end{figure}

Although the dominant decay mode of $Z_H$ is into dijets, the discovery 
modes for $Z_H$ would be $l^+l^-$ (with a branching 
ratio of 2.5\% for $e^+e^-$, $\mu^+\mu^-$ and $\tau^+\tau^-$ individually). 
%and 
%$T\bar{T}$ (with a branching ratio of 2-7\%), 
%$t\bar{t}$ (with a branching ratio of 2-5\%).  
The di-lepton mode $e^+e^-$ or $\mu^+\mu^-$ provides a clean signal, which 
can be separated from the SM background by studying
the invariant dilepton mass distribution, as shown in Fig.~\ref{fig:ZHeE}.
Searches for heavy neutral gauge boson in dilepton final states have been studied at 
both the Tevatron \cite{dileptonZH, RunIINP, RunIIZH} and the LHC \cite{atlasTDR, dileptonZH, LHCZH}.  
The current search limit from
the Tevatron Run II is about 600$-$900 GeV \cite{RunIINP, RunIIZH}, 
while mass up to about 5 TeV 
could be covered at the LHC \cite{atlasTDR, LHCZH}.

The pair production of $t\bar{t}$  (with a branching ratio of 2-5\%) via $Z_H$ decay can also be useful.  
Searches of  $t\bar{t}$ resonance have been
studied in \cite{atlasTDR, wbwbLHC}. Requiring that one $W$ decays leptonically and 
one $W$ decays hadronically, the signal is $l\nu bbjj$.
The dominant backgrounds are $W+$jets, $Z+$jets,
$t\bar{t}$ and $tbj$.
Requiring large missing $E_T$, energetic isolated electron or muon, at least
four energetic jets with at least one  tagged as a $b$-jet reduces some of 
the backgrounds.  The reconstruction of the $t\bar{t}$ resonance mass could be used to
further suppress the continuum $t\bar{t}$ background.  
For 300 ${\rm fb}^{-1}$
integrated luminosity, $5\sigma$ discovery limit of 
$\sigma\times {\rm Br}$ are 835 fb, 265 fb and 50 fb for the masses of resonances being 
$m=$ 500 GeV, 1 TeV and 2 TeV \cite{atlasTDR}.  
With the cross section and $t\bar{t}$ branching ratio of $Z_H$, the reach 
for $Z_H\rightarrow t\bar{t}$ at the LHC is only about 1 TeV, due to the small decay branching 
ratio into $t\bar{t}$ final states.

It is also possible to discover the heavy $Z_H$ gauge boson via its decaying 
into a pair of heavy top quarks $T\bar{T}$,  with a branching ratio of 2-7\%.  The heavy top mostly decays into 
$\phi^\pm b$, which typically has two $W$'s and six $b$-jets in the final states.  
Such channel, however, also suffers from small $Z_H\rightarrow T\bar{T}$ branching ratio,
and its LHC reach is limited.

%Requiring one $W$ decay leptonically
%and the other $W$ decay hadronically, the final states consist of  at least 
%two b-jets $+$ lepton $+$ missing $E_T$.  The SM background of $t\bar{t}$ can 
%be suppressed by requiring at least one $T$ decay into $\phi^\pm b$ or 
%$Z t$, which leads to two more b jets or dilepton that are absent in 
%$t\bar{t}$ events.  
  
\subsection{Higgses}

\subsubsection{SM Higgs}
The SM Higgs mass can be obtained via the minimization of the Higgs potential,
which depends on $f(\hat{f})$, $M$ and $\Lambda$.  Varying $M$ between 0 and 150 GeV,
$\Lambda$ between $2 \pi f$ and $4 \pi f$, and $f$ between 500 GeV and 
1500 GeV, the Higgs mass is found to be in the 
range of $145-180$ GeV.  For this 
intermediate mass region, several channels have been studied for Higgs 
discovery.

The best channel for Higgs discovery at the LHC for intermediate mass region is vector boson fusion production, 
with $h\rightarrow WW^*\rightarrow l\nu l\nu$  \cite{qqH, WBFWW}.
Signals for such channel are two forward tagging jets, central jet veto, energetic di-leptons and 
missing $E_T$ from neutrinos.  
The characteristic signatures of additional forward jets in the detector and low jet activity in the central region allow for an efficient background rejection.   
The remaining dominant backgrounds come from $t\bar{t}$ and QCD $\gamma^*/Z+$jets production
with $\gamma^*/Z\rightarrow ll$.  Requiring a tag forward jet not being tagged as $b$-jet reduces the $t\bar{t}$ background. $ee$ and $\mu\mu$ Drell-Yan backgrounds can be efficiently rejected by 
tightening the di-lepton mass cut and by introducing a $E_T^{miss}$ cut.   Analyses  in 
Ref.~\cite{qqH} showed that such process has a better signal-to-background ratio than $gg\rightarrow h$ with $h\rightarrow WW^*$ or $ZZ^*$ for Higgs mass between 140 GeV and 190 GeV.
At ATLAS, a sensitivity of  5 $\sigma$  can be reached with  an integrated luminosity of only 10 ${\rm fb}^{-1}$ in such channel.

Gluon fusion process $gg\rightarrow h$ has the largest cross section
for Higgs production at the LHC.  For the intermediate mass region, 
the so-called golden plated channel $h\rightarrow ZZ^*\rightarrow 4l$
made of 4$e$, 4$\mu$ and 2$e$2$\mu$ decays, provides a clean signature.  
The most important irreducible backgrounds are $ZZ^*$ and $Z\gamma^*$ production
with decays to four leptons.  The most important reducible backgrounds 
are $t\bar{t}$ and $Zb\bar{b}$ production.  The main cuts to reduce the 
background are isolated leptons, a mass cut on one of the lepton pairs to be around 
the $Z$ mass, and a requirement for the other lepton pair to have an invariant 
mass above 20 GeV\cite{HZZ}.  It is shown that with an integrated luminosity of 
30 ${\rm fb}^{-1}$, this channel may allow discovery above 5 $\sigma$ in the 
range of $130<m_{h}<180$ GeV \cite{atlasTDR}, with an exception near 170 GeV, 
where this branching ratio is reduced due to the 
opening of $h\rightarrow WW$ 
decay.

In the region around 170 GeV, we can use $h\rightarrow W W^* \rightarrow 
l \nu l\nu$ channel.  The irreducible backgrounds are made of $WW$ continuum, 
and of $WZ$ and $ZZ$.  The reducible backgrounds come from $t\bar{t}$, $Wt$,
$Wbb$, $b\bar{b}$ and $W+$ jet production.  Requesting central jet veto, strong
angular correlation between the leptons and high missing transverse mass 
allows us to discriminate between the signal and the background.  With an integrated 
luminosity of 10 ${\rm fb}^{-1}$, a significance larger than 5 $\sigma$ maybe 
obtained in the region $150 < m_{h} < 190$ GeV~\cite{HWW}.

\subsubsection{$\phi^0$ and $\phi^\pm$}

Besides the SM Higgs, there are three additional Higgses that couple 
to both the SM fermions and the gauge bosons: one neutral Higgs $\phi^0$ and
a pair of charged Higgses $\phi^\pm$.

The light neutral Higgs boson $\phi^0$ is a pseudo-scalar and charged under the spontaneously broken ${\rm SU}(2)_R$. Its mass is a free parameter and is determined by 
$\mu_r$ that can be anything below $f$. $\phi^0$ can in principle be very heavy and become unobservable at the  LHC. Here we consider another possibility where the mass of $\phi^0$ is 
about 100 GeV. Due to its pseudo-scalar nature, 
there is no $\phi^0W^+W^-$, $\phi^0ZZ$ coupling at tree 
level \footnote{$\phi^0\phi^0WW$ coupling, however, is allowed at tree-level.
Its coefficient depends on the choice of the Higgs non-linear representation.  
For our choice of Higgs representation as 
in Eq.~(\ref{eq:Higgsrep}), $\phi^0\phi^0WW$ coupling is non-zero.  However, if a non-linear representation of the Higgs field similar to those defined in Ref.~\cite{rainwater} 
is used, $\phi^0\phi^0WW$ coupling is zero. 
Any physical observable, however, does not depend on the choice of the Higgs representation.}.
Such couplings, similar to $\phi^0\gamma\gamma$ and $\phi^0 g g$, can be generated at loop level with heavy fermions.
%the $WW$, $ZZ$, $WW^*$ and $ZZ^*$ decay modes and  
%the Higgsstraahlung processes $pp\rightarrow W^\pm \phi^0, Z \phi^0 $ are  
%suppressed by at least a loop factor. 

$\phi^0$ decays donimantly into $b\bar{b}$, $c\bar{c}$ or
$\tau^+\tau^-$.  The decay widths are proportional to the square 
of the corresponding Yukawa couplings, with an additional suppression factor of $v^2/(2f^2)$ comparing to 
that of the SM Higgs.  The decay branching ratio of 
$\phi^0\rightarrow b\bar{b}$, $c\bar{c}$ and $\tau^+\tau^-$, however, are close to the corresponding SM Higgs decay branching ratios, since the additional suppression factor cancels out. 
Given the huge QCD background in
the LHC environment, the discovery of $\phi^0$ is difficult through those channels, unless there are
other particles produced associated with $\phi^0$, which could
provide a handle to trigger the events and  to distinguish the background \cite{rainwater}.

Similar to the SM Higgs, the loop generated $\phi^0\rightarrow \gamma\gamma$ 
could be useful due to the narrow 
$\gamma\gamma$ peak that can be reconstructed to distinguish the signal from the background. 
Unlike the SM Higgs, where $h\gamma\gamma$ are generated by both the top quark and $W$ loop, the one-loop SM gauge boson contribution to $\phi^0\gamma\gamma$ is zero because of the absence of the tree level $\phi^0WW$ coupling.  The SM top loop contribution is also suppressed since $\phi^0 t \bar{t}$ coupling is suppressed by small $M/f$.  $\phi^0\gamma\gamma$ coupling, however, gets contributions from the loop with the heavy top partner $T$, with an unsuppressed $\phi^0T\bar{T}$ coupling.  
Due to the heavy top mass,  heavy top quark loop contribution to $\phi^0\gamma\gamma$ is suppressed by a factor of $v/(\sqrt{2}f) $ comparing to the SM top contribution to $h\gamma\gamma$ for 
$m_h=m_{\phi^0}$.  The heavy gauge boson loop contributions are absent since there is no $\phi^0W_HW_H$ coupling.
Since the SM top loop competes with the SM $W$ loop in its contribution to $h\gamma\gamma$, 
the decay width of $\phi^0\rightarrow \gamma\gamma$ is roughly $v^2/(2f^2)$ suppressed comparing to the decay width of $h\rightarrow \gamma\gamma$.  Given that $\phi^0\rightarrow b\bar{b}$ is also suppressed by the same factor comparing to $h\rightarrow b\bar{b}$,   the branching ratio of $\phi^0\rightarrow 
\gamma\gamma$ is roughly the same as ${\rm Br}(h\rightarrow\gamma\gamma$) for
$m_h=m_{\phi^0}$.

The associated production of $\phi^0$ with $W$ or $Z$ is suppressed by a loop factor comparing to the usual Higgsstrahlung production at the LHC, due to the absence of the tree level $\phi^0WW$ and $\phi^0ZZ$ coupling.  The dominant production is again the gluon fusion process $gg\rightarrow \phi^0$ with a heavy top loop.  Similar to $\phi^0\gamma\gamma$ coupling that discussed above, gluon fusion production of $\phi^0$ is suppressed by a factor of $v^2/(2f^2)$ comparing to that of the SM Higgs with the same mass. 
 The total number of event of $gg\rightarrow \phi^0\rightarrow \gamma\gamma$ is then suppressed by a factor of $v^2/(2f^2)$  comparing to the SM process $gg\rightarrow h \rightarrow  \gamma\gamma$.
Studies for the SM Higgs discovery in this channel \cite{atlasTDR, CMSHiggs} showed that a 5$\sigma$ discovery of a  light (115 GeV) SM Higgs requires an integrated luminosity of about 
25 ${\rm fb}^{-1}$ at the LHC.  Since the significance level scales as  $(\sigma\times {\rm Br})_{\rm signal}\times \sqrt{\cal L}$,
a factor of 9 suppression of the $\phi^0$ signal cross section (for a low value of $f\sim$ 500 GeV) is very hard to 
compensate with an increasing luminosity.

\begin{figure}
\resizebox{4 in}{!}{
\includegraphics{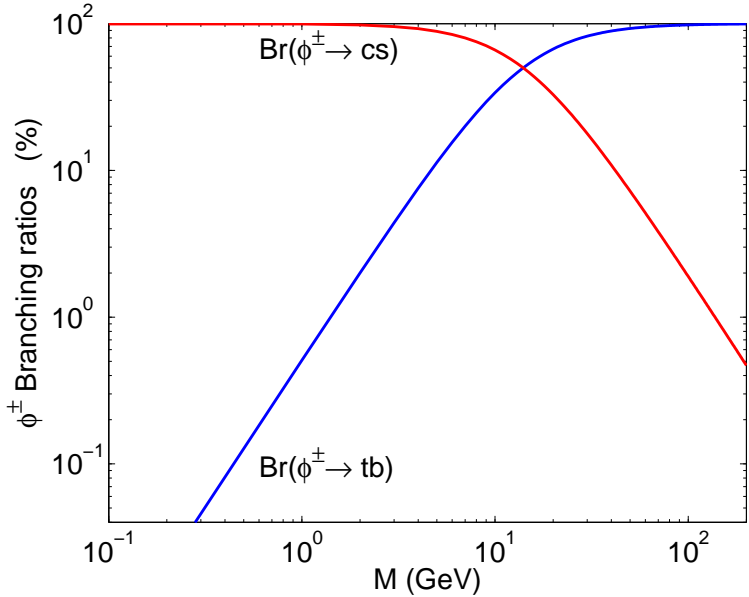}
}
\caption{The decay branching ratios of $\phi^\pm\rightarrow tb$ and 
$\phi^\pm \rightarrow cs$, as a function of $M$.
The model parameters are chosen to be 
$f=800$ GeV, $\mu_r=50$ GeV, and $\Lambda=4 \pi f$.  
 }
\label{fig:hRdecay}
\end{figure}

The charged Higgses $\phi^\pm$ dominantly decay into $tb$ or $cs$, with the 
decay width of the former channel proportional to $(M/f)^2$, and the decay width of 
the latter channel proportional to the charm Yukawa coupling squared.
Fig.~\ref{fig:hRdecay} shows the branching ratios of $\phi^\pm\rightarrow tb$ and 
$\phi^\pm \rightarrow cs$ as a function of $M$.
It is clear that for larger value of $M$, $\phi^\pm\rightarrow tb$ dominates.
If the particles produced associated with $\phi^\pm$ do not involve 
leptons, the $W$ from top decay
is required to decay leptonically, which can be used as a trigger, and also 
to suppress the background.  
For very small value of $M$ $\lesssim$ 1 GeV, 
${\rm Br}(\phi^\pm\rightarrow tb)$ drops to less than 1\% and $\phi^\pm\rightarrow cs$
dominates, which leads 
to completely different phenomenology.
We defer the discussion of such case together with $M=0$ limit 
to  Sec.\ref{sec:M0}.

\begin{figure}
\resizebox{6.3 in}{!}{
\includegraphics{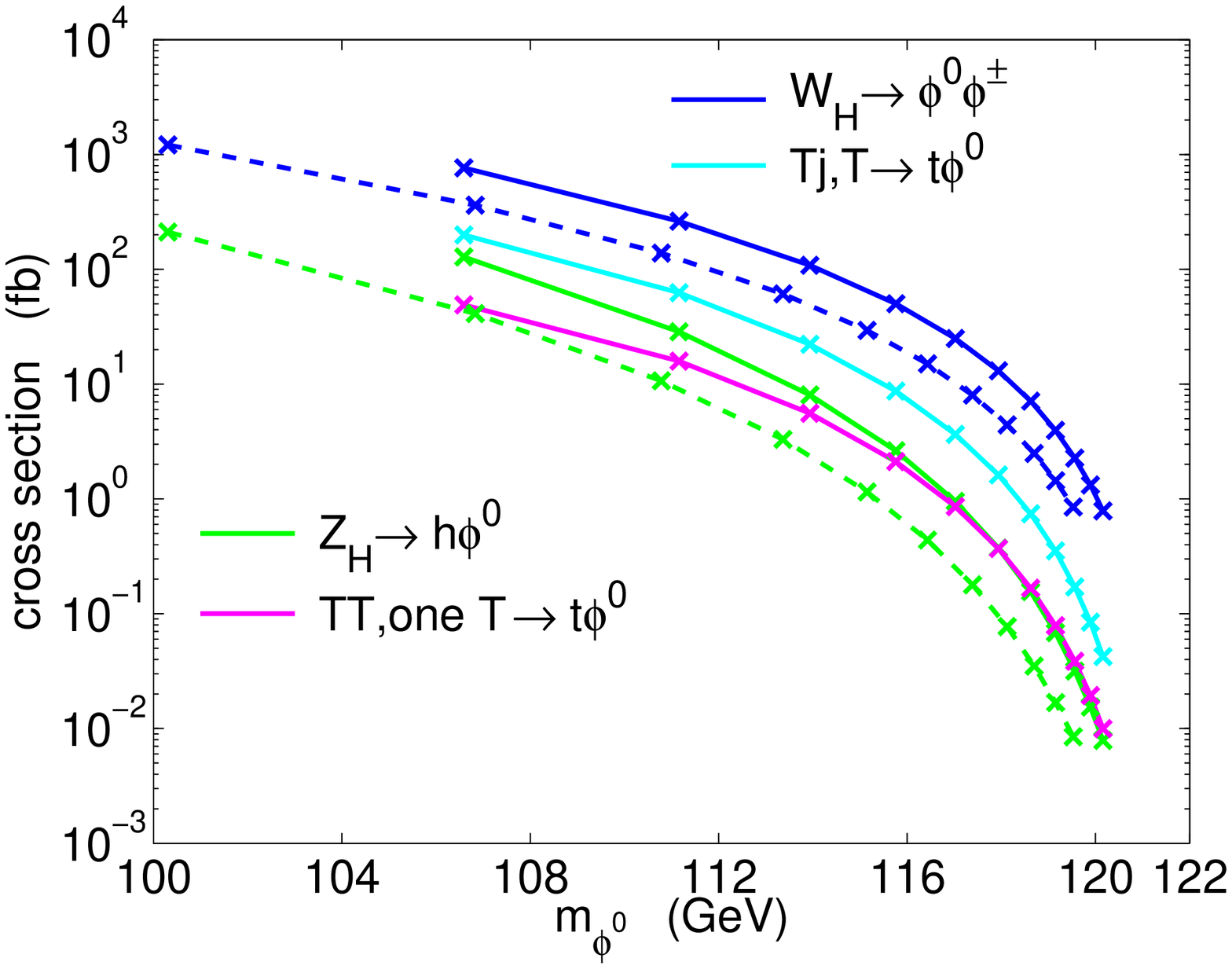}
\includegraphics{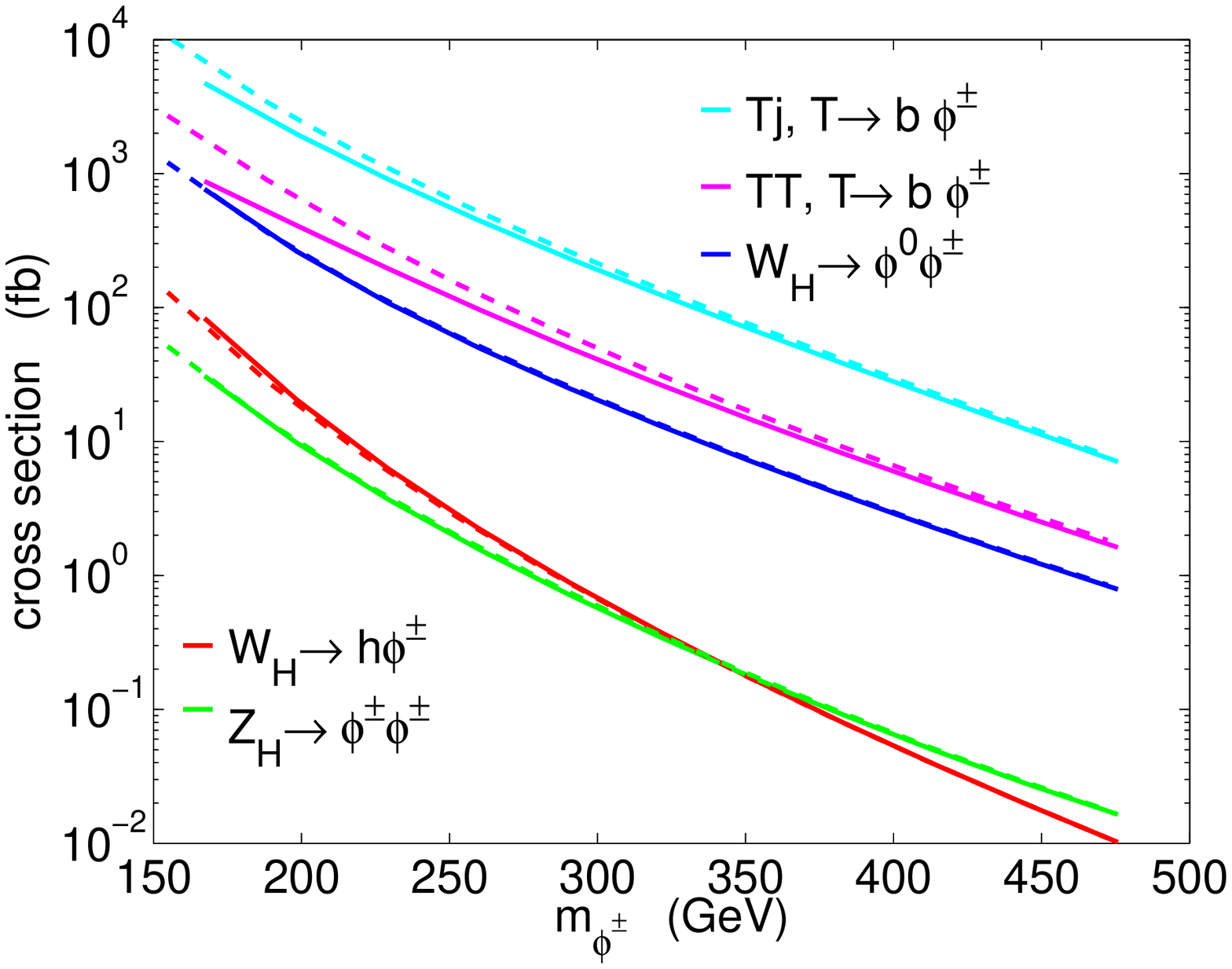}}
\caption{The production of $\phi^0$ (left plot) 
and $\phi^\pm$ (right plot) from
the decay of heavy particles at the LHC.
The solid lines are for $M=150$ GeV, while the dashed lines are for 
$M=0$.
In the left plot, from top to bottom, the production modes are 
$W_H\rightarrow \phi^0\phi^\pm$, single $T$ production with $T\rightarrow t\phi^0$,
$Z_H\rightarrow h\phi^0$ and  $TT$ pair production with
one heavy top decaying into $\phi^0$ while the other top 
decaying into anything.  
In the right plot, from top to bottom (for small $m_{\phi^\pm}$), the production modes are
single $T$ production with $T\rightarrow b\phi^\pm$,
$TT$ pair production with
both heavy tops decaying into $ b \phi^\pm$,
$W_H\rightarrow \phi^0\phi^\pm$, $W_H\rightarrow h\phi^\pm$,
and $Z_H\rightarrow \phi^+\phi^-$.
The other parameters are chosen to be $\Lambda=4 \pi f$ and $\mu_r=50$ GeV.
%, and an additional 
%factor of two has been multiplied.
%For $Z_H \rightarrow \phi^\pm\phi^\mp$, we already
%multiply a factor of two to take into account the fact that two $\phi^\pm$ 
%can be produced from $Z_H$ decay.
The ``x'''s in the left plot correspond to the 
value of $f$ being $500,600,...,1500$ GeV.
}
\label{fig:phi0phipm}
\end{figure}

The heavy particles in the LRTH models, $W_H$, $Z_H$, and 
$T$, can decay into the light Higgses.  Due to the large Drell-Yan cross 
sections for $W_H$ and $Z_H$, and the large single $T$ production cross 
section at the LHC, the  
production of $\phi^0$ and $\phi^\pm$ from the decay of heavy particles
could be sizable,  
as shown in Fig.~\ref{fig:phi0phipm}.   Notice that the 
fall of the cross section  for heavier Higgs mass is due to 
the reduction of $W_H$, $Z_H$ and $T$ production cross sections
with increasing $f$.

For the neutral Higgs $\phi^0$, 
the dominant production mode is through $W_H \rightarrow \phi^0\phi^\pm$,
with a cross section of about $10^3$ fb $-$ 1 fb.  Combined with the 
decay of $\phi^0$ and $\phi^\pm$, we can look for signals of 
4 $b$-jets $+$ 1 lepton ($e$ or $\mu$)  $+$ missing $E_T$.  Two $b$-jets need 
to be chosen to reconstruct the $\phi^0$ mass, while $\phi^\pm$ 
can be reconstructed as described above.
$\phi^0$ produced from the heavy top decay: $T\rightarrow t \phi^0$, 
might also be used to identify the neutral Higgs.

For the charged Higgses $\phi^\pm$, the dominant production mode is through heavy
top decay, since the branching ratio for $T\rightarrow \phi^\pm b$ is more than 
70\%.  
The cross section is in the range of $6\times 10^3$ fb $-$ 10 fb.
Considering the single heavy top production $pp\rightarrow T jX$,
with $T \rightarrow \phi^\pm b$, the signal is 
3 $b$-jets $+$ 1 jet $+$ 1 lepton ($e$ or $\mu$) $+$ missing $E_T$.  The top quark 
from $\phi^\pm$ decay can be reconstructed through $bW$, while $\phi^\pm$ can
be reconstructed through $tb$.  The reconstructed invariant mass 
for $tb$ could also tell us the mass of $\phi^\pm$. 

$\phi^0$ and $\phi^\pm$ can also be produced in association with the third 
generation quarks: 
$b\bar{b}\phi^{0,\pm}$, 
$t\bar{t}\phi^{0,\pm}$, $tb\phi^{0,\pm}$.  The cross sections are
usually much smaller than the ones that are mentioned above, and
therefore are not discussed further.

\subsubsection{$\hat{h}_1^\pm$ and $\hat{h}_2^0$ }

The complex charged  and neutral Higgses $\hat{h}_1^\pm$ and $\hat{h}_2^0$ 
couple to the gauge bosons only.  Their masses are very degenerate, with 
a small mass splitting of about 100 $-$ 700 MeV 
introduced by the electro-magnetic interactions.  The 
charged Higgses $\hat{h}_1^\pm$ are slightly heavier than the neutral one, and 
can therefore decay into $\hat{h}_2^0$ plus soft jets or leptons.  If 
the decay happens inside the detector, the jets
and leptons are so soft that they  cannot be detected at colliders.  
The neutral Higgs 
$\hat{h}_2^0$ is stable and escapes the detector, and therefore
appears as a missing energy signal.  
It is, however, a good dark matter
candidate.  The study of $\hat{h}_2^0$ as a viable dark matter candidate is 
left to future studies \cite{suDM}.

The production cross sections of $\hat{h}_1^\pm$ and $\hat{h}_2^0$ at the LHC are relatively 
small.  They can only be pair produced via the exchange of 
photon, $Z$, $W$, $Z_H$, or Higgses. 
The cross sections are about 1 fb.  
The collider signatures depend on the lifetime of $\hat{h}_1^\pm$, 
which further depends on the mass splitting $\Delta{M}$ 
between $\hat{h}_1^\pm$ and $\hat{h}_2^0$.
For small $\Delta{M}<m_\pi$, the decay lifetime of $\hat{h}_1^\pm$ is relatively long and the 
decay of $\hat{h}_1^\pm\rightarrow \hat{h}_2^0$ happens outside the 
detector.  $\hat{h}_1^\pm$
appears as a charged track in the detector with little hadronic activities. 
It can be distinguished from the muon background by requiring a large 
ionization rate $dE/dx$ or using the time of flight information. 
Such signal is hard to miss since it is almost background free.
For an extensive review on the collider searches of  a long lived stable particle, see Ref.~\cite{stable}.
For $\Delta{M}\sim m_\pi$, 
$\hat{h}_1^\pm$ decay inside the detector while leaving a track in the 
tracking chamber, such events could be identified with a disappearing track.
To trigger on such events, we need to look at the associated production of 
$\hat{h}_1^\pm$ with a jet.
For larger $\Delta{M}$, $\hat{h}_1^\pm$ decay instantly inside the detector, 
the soft jets and leptons escape the detection, and 
the missing $E_T$ is balanced in the pair production.
Such events are very difficult to detect since  there is no 
visible final states to be observed.  Similar studies for degenerate winos
in the anomaly-mediated supersymmetry breaking scenario have been 
done in the literature~\cite{wino}.

\section{Collider phenomenology with $M=0$ or very small value of $M$}
\label{sec:M0}
All the above discussions are for small but sizable value of $M$. 
From Eq.~(\ref{eq:topmix}), the top quark mass eigenstates $t$ and $T$ are 
related to the gauge eigenstates $u_{3L}, u_{3R}, q_L, q_R$ 
by the mixing angle 
$\alpha_L$ and $\alpha_R$.  In the limit of $M=0$, $\sin\alpha_L=0$ and 
$\sin\alpha_R=0$.  Therefore, the SM top quark is purely $(u_{3L}, q_R)$, 
and the
heavy top is purely $(q_L, u_{3R})$.   Certain couplings vanish at this limit,
as shown in Table~\ref{tab:M0}.  
The couplings that vanish at $M=0$ are proportional to $M/f$.  
We will discuss below the collider phenomenology of $M=0$ case.  They can also be applied to the case when 
$M$ deviates from zero slightly: $M\lesssim 1$ GeV.

The main phenomenological difference between the $M=0$ case and  the case 
discussed in the previous section comes from the 
decay modes of $\phi^\pm$.
Because of the absence of $\phi^+ \bar{t}b$ coupling, $\phi^\pm$ can no longer
decay into $tb$.  
$\phi^\pm$ cannot
decay into ${T}b$ either since $m_T>m_{\phi^{\pm}}$.
The previous subdominant channel $\phi^\pm\rightarrow cs$
now becomes the main decay mode, leading to all jet final states.
For nonzero value of $M$, Fig.~\ref{fig:hRdecay} shows that $\phi^\pm\rightarrow cs$ becomes dominant
(${\rm Br}(\phi^\pm\rightarrow tb)<1\%$) when $M\lesssim 1$ GeV.
%The only possible decay channels are $\phi^\pm\rightarrow
%\phi^0 q\bar{q}^\prime$ or
% $\phi^\pm\rightarrow h q\bar{q}^\prime$ via 
%an off-shell $W_H$, as shown in 
%Fig.~\ref{fig:phipmdecay}. Most of $\phi^\pm$ decays into 
%$\phi^0 q\bar{q}^\prime$, leading to $bbqq^\prime$ or 
%$\tau\tau qq^\prime$ final states.  
The discovery of $\phi^\pm$ becomes
extremely difficult at the LHC, due to the huge QCD jet background.

%Less 1\% ({\bf check the number})
%of $\phi^\pm$ decays into $h q\bar{q}^\prime$, with $h$ 
%subsequently decays into $WW^{(*)}$.  $\phi^\pm$ might be discovered through 
%this channel if at least one of the $W$ decay leptonically.  However, the 
%numbers of signals is suppressed due to the small branching ratio.

Due to the absence of certain couplings in  the $M=0$ limit, some production
processes for $\phi^0$ disappear.
The cross sections for $\phi^0$ from the decay of heavy particles 
for $M=0$ are given in the dashed lines of the left plot of 
Fig.~\ref{fig:phi0phipm}.
No contribution from $T$ decay is present since $T t \phi^0$ coupling is zero.
For the same $m_{\phi^0}$, the cross section for $M=0$ is smaller than 
non-zero $M$ case.  However, when we compare the cross section with the 
same value of $f$, the one for $M=0$ is actually larger.  This is because
$\hat{f}$ is smaller for $M=0$ case, which leads to a smaller 
mass for the heavy gauge boson $W_H$ and $Z_H$ and  a larger 
Drell-Yan cross section.  
The decay of $\phi^0$ is still the same as before: 
$\phi^0\rightarrow b\bar{b}, c\bar{c}, \tau\bar{\tau}$.  
$\phi^0$ is dominantly produced 
associated with $\phi^\pm$ from $W_H$ decay.
This channel is not so useful for $\phi^0$ discovery at the LHC
since both $\phi^0$ and $\phi^\pm$ decay hadronically.  The cross 
section for $\phi^0$ produced associated with a SM Higgs from $Z_H$ decay 
is about a factor
of 10 smaller than $\phi^0\phi^\pm$ production.  The leptonic final states
from Higgs decay might make this channel useful for $\phi^0$ discovery.

The cross sections for $\phi^\pm$ production from heavy particle decays for $M=0$  are
presented in the dashed lines in the right plot of Fig.~\ref{fig:phi0phipm}.
For $\phi^\pm$ production from heavy top decay, the cross section
is larger than the non-zero $M$ case, this is mainly because the branching
ratio for $T\rightarrow \phi^+ b$ is larger, now 100\%.  The discovery 
of $\phi^\pm$, however, is very difficult, because $\phi^\pm$ dominantly decay
hadronically.  The suppressed 
$h\phi^\pm$ production from $W_H$ decay might become important for 
$\phi^\pm$ studies.

For the heavy top, both
the single and pair heavy top production cross sections do not change much.
However,  the heavy top decay is affected.  The only two body 
decay mode is now $T\rightarrow b \phi^+$, with a branching ratio of 100\%.
The other decay channels: $T\rightarrow b W^+$, $T\rightarrow t Z$,
$T\rightarrow t \phi^0$ and $T\rightarrow t h$ are forbidden
since the relevant couplings are zero.
Due to the dominant hadronic decay of $\phi^\pm$ for $M=0$, the discovery
of the heavy top quark also becomes difficult at the LHC. 

The situation is different for the $W_H$ and $Z_H$.
The Drell-Yan cross section for $W_H$ and $Z_H$ do not change much since 
they only depend on the masses of the heavy gauge bosons. 
The decays of $W_H$ and $Z_H$ almost do not change, except that 
${\rm Br}(W_H\rightarrow tb)={\rm Br}(Z_H\rightarrow T t)=0$.  
These two branching 
ratios are small for non-zero $M$ (less than a few percent).  Shutting off 
these two decay modes does  not change the branching ratio of other decay 
channels that much.  The dilepton signal and $t\bar{t}$ signal for $Z_H$ 
do not change.  For $W_H$, its discovery potential depends on the 
masses of $\nu_R$.  If $m_{\nu_R}<m_{W_H}$, $W_H$ can be studied using 
dilepton plus jets signal from $W_H\rightarrow l \nu_R$ process.
If $m_{\nu_R}>m_{W_H}$, however, $W_H$ discovery also becomes challenge at the LHC.
The study of its decay to $Tb$
is very hard due to the difficulty of identifying $T$, as discussed above.
Signals suffer from either huge QCD background or small cross sections
for processes with leptonic final states.

\section{Conclusion}
\label{sec:conclusion}

The twin Higgs mechanism provides an alternative method to solve the little 
hierarchy problem.  In this paper, we present in detail the embedding of 
the twin Higgs mechanism in LRTH models.  There are TeV scale 
heavy  top and heavy gauge bosons, which interact with SM quarks, 
leptons  and gauge bosons.  There are also additional Higgses in the model.
The neutral Higgs $\phi^0$ and charged Higgses $\phi^\pm$ couple to the 
SM quarks, leptons and gauge bosons.  There is an extra ${\rm SU}(2)_L$ Higgs
doublet $(\hat{h}_1^+, \hat{h}_2^0)$, which couples to the gauge sector only. 
The lighter one $\hat{h}_2^0$ is stable, which could be a good dark matter 
candidate.

The collider phenomenology of the LRTH depends sensitively  on the parameter $M$, 
which is 
the mass mixing between the vector heavy top singlet.  The discovery potential
at the LHC for $M\gtrsim 5$ GeV is very promising.
For the heavy top, the dominant production channel at the LHC is single heavy
top production in association with a jet.  Heavy top dominantly decays to
$\phi^\pm b$.  The consequent decay of $\phi^\pm$ leads to signals of 
$l \nu bbbj$.  The reconstruction of the intermediate on shell particles 
could distinguish the signal from the background.  $W_H$ and $Z_H$ 
are produced via the Drell-Yan processes.  If $\nu_R$ is too heavy for
$W_H$ to decay into, $W_H$ could be discovered via $T b$ or $tb$ channel.
If $m_{\nu_R}<m_{W_H}$, $W_H\rightarrow l \nu_R$ could also be used to identify
$W_H$.
The dilepton decay mode for $Z_H$ provides a clean signal, although $Z_H$
could also be studied in $t\bar{t}$ or $T \bar{T}$ channel.

The mass of the SM Higgs is in the range of 145$-$180 GeV. Its 
discovery via $ZZ^*$ or $WW^*$ is promising at the LHC.  The charged Higgses
$\phi^\pm$ and the neutral Higgs $\phi^0$ are most likely to be discovered
in the decay products of heavy particles.  The charged Higgses $\phi^\pm$, which are largely
produced in $T$ decay, decays dominantly to $tb$.  The discovery for 
$\phi^0$ is much more difficult.  It can be produced from 
$W_H\rightarrow\phi^0\phi^\pm$, and decays dominantly into $b\bar{b}$.  

The Higgses $\hat{h}_1^\pm$ and $\hat{h}_2^0$ can only be pair 
produced via electroweak processes at the LHC.  Their masses 
are very degenerate, and $\hat{h}_1^\pm$ decay to 
$\hat{h}_2^0$ plus soft leptons or jets.  The collider signatures depend strongly
on the mass splitting $\Delta{M}$ between $\hat{h}_1^\pm$
and $\hat{h}_2^0$.
If $\Delta{M}\lesssim m_\pi$, the decay lifetime of $\hat{h}_1^\pm$
is relatively long.  We will see either isolated track in the tracking chamber with 
little hadronic activities, or disappearing tracks.  Otherwise, both the soft 
jets or leptons, and the missing energy from $\hat{h}_2^0$  escape the 
detection.  It becomes difficult to identify  $\hat{h}_1^\pm$ and $\hat{h}_2^0$
at the LHC.  The stable  $\hat{h}_2^0$ could be a good dark matter candidate.
Its relic density analysis and the direct and indirect detection potential
are under current investigation \cite{suDM}.

If the mixing $M$ between the vector top singlet is  very small $\lesssim 1$ GeV, 
the mixings between the two top quark gauge eigenstates are negligible.  
Certain couplings, for 
example, $\phi^\pm tb$, go to zero, which leads to dramatic changes in the 
collider phenomenology.  Most of the signals discussed for sizable $M$
suffer from either huge
QCD jet background, or small cross sections for signals with leptonic 
final states.  The only exceptions are $W_H$ (if $m_{\nu_R}<m_{W_H}$) and
$Z_H$, which can still be discovered via Drell-Yan production and 
their leptonic  decays.

There are further studies can be performed in the LRTH model.  In this paper,
we analyze the productions of new particles and the general feature of
their decay patterns.  
A more realistic analysis would 
include both the signal and the background, and the choices of 
appropriate cuts to either trigger the events, and/or to 
suppress the background.  Therefore, it is worthwhile to 
pick typical decay processes and study in detail the LHC reach 
of the LRTH model.
For example, for heavy top, the dominant
production mode is single heavy top production $pp\rightarrow T j X$,
with the subsequent decay of 
$
T \rightarrow \phi^+ b,\ \ \phi^+ \rightarrow t \bar{b}, \ \ 
t \rightarrow W^+ b \rightarrow l^+ \nu b.
$
The collider signal is three $b$-jets $+$ one jet $+$  one lepton  $+$ 
missing $E_T$.  
More than 10,000 events
can be seen at $10\ {\rm fb}^{-1}$ luminosity
for a heavy top of around 600 GeV. 
Detailed study need to be done to optimize the cuts and identify the signal
from the background \cite{collidersu}.  

It is also important to identify,
experimentally, the twin Higgs mechanism.  In particular, the equality of 
the left and right Yukawa couplings.  A careful examination of the cancellation
between the quadratically divergent 
contributions from SM-like light  top and heavy top quark shows that 
the following leading order relation needs to be satisfied: 
\begin{equation}
y_L^2 - \frac{y_R}{f}m_{T}=0
\label{eq:testTH}
\end{equation} 
Therefore,
to identify the twin Higgs mechanism, it is essential 
to testify this  relation at colliders. 
The left Yukawa coupling $y_L$ could be obtained from the SM top quark
mass $y_L = \sqrt{2} m_t / v$.  
The mass of the heavy top $m_{T}$ can be reconstructed 
from the heavy top decay chain.  
%In the limit of 
%small mixing between the singlet vector top pair,  
%the decay width of the heavy top 
%is dominated by $T \rightarrow \phi^+ b$:
%\begin{equation}
%\Gamma(T \rightarrow \phi^+ b) \approx \frac{y_R^2}{64 \pi} m_{T}.
%\end{equation}
Knowing $m_T$, the right Yukawa coupling $y_R$ can be obtained from the heavy top decay width
$\Gamma(T\rightarrow \phi^+ b)$ using Eq.~(\ref{eq:tHdecay}).
The value of $f$ could be derived from $m_T$ and $y_R$ using the relation that 
$f=m_T/y_R$.
Studies on testifying the twin Higgs mechanism  along this direction are
under current investigation \cite{mechanismsu}.

The collider signatures of the LRTH model could mimic 
signals of the little Higgs models.  Both classes of models have similar particle 
content: heavy top and heavy 
gauge bosons.   If we 
see heavy top and heavy gauge bosons at collider, it is important to identify
whether they are the ones from the LRTH, or the ones 
from the little Higgs models.   There are several handles that we can use to 
distinguish these two models, for example, the mass relation between
heavy top and heavy gauge bosons, and the decay pattern of the heavy top quark.
The Higgs sector of the 
LRTH might also mimic that of two Higgs doublet models.
Further studies are needed to distinguish those scenarios.

\begin{acknowledgments}
We would like to thank Z. Chacko for useful discussion on the twin Higgs
model.  We also would like to thank T. Han and L. Wang for 
discussion of collider signals, E. Dolle for cross checking  the model 
files for CalcHEP, and A. Pukhov for help with CalcHEP.  
We  thank the referee for careful reading of the draft and useful comments and suggestions.
This work is supported under U.S. Department of Energy
contract\# DE-FG02-04ER-41298.

\end{acknowledgments}

%\begin{appendix}
\appendix

\section{Higgs fields in unitary gauge}
\label{app:unitary}
The scalar fields of the nonlinear sigma model can be parameterized by
\begin{equation}
    H = f e^{i\frac{\pi}{f}}\left(%
\begin{array}{c}
  0  \\
  0  \\
  0  \\
  1  \\
\end{array}%
\right),\ \ \ \ 
    \pi= \left(%
\begin{array}{cccc}
  -N/2 & 0 & 0 & h_1 \\
  0 & -N/2 & 0 & h_2 \\
  0 & 0 &  -N/2 & C \\
  h_1^* & h_2^* & C^* & 3N/2  \\
\end{array}%
\right),
\end{equation}
where $\pi$ are the corresponding Goldstone fields. 
$N$ is a neutral real pseudoscalar, $C$ and $C^*$
is a pair of charged complex scalar fields, and $(h_1,h_2)$ is the
SM ${\rm SU}(2)_L$ Higgs doublet. They together comprise the seven
Goldstone bosons.  Similar expression can be written down for Higgs
field $\hat{H}$ with Goldstone fields $\hat{\pi}$.

Re-summing the
exponential expansions, these Goldstone boson fields can be
parameterized by
\begin{equation}
\label{GBrep}
    H= i\frac{\sin\sqrt{\chi}}{\sqrt{\chi}}e^{i\frac{N}{2f}}\left(%
\begin{array}{c}
  h_1 \\ h_2\\
  C\\
  N-if\sqrt{\chi}\cot\sqrt{\chi}\\
\end{array}%
\right),
\ \ 
\hat{H} = i\frac{\sin\sqrt{\hat{\chi}}}{\sqrt{\hat{\chi}}}
e^{i\frac{\hat{N}}{2\hat{f}}}\left(%
\begin{array}{c}
  \hat{h}_1 \\ \hat{h}_2\\
  \hat{C}\\
  \hat{N}-i\hat{f}\sqrt{\hat{\chi}}\cot\sqrt{\hat{\chi}}\\
\end{array}%
\right),
\end{equation}
where $\chi = (h_1^{\dagger}h_1+h_2^{\dagger}h_2+C^*C+N^2)/f^2$
and similarly for $\hat{\chi}$. It can be shown explicitly that
this parametrization has a canonically normalized kinetic term for
every Goldstone field except $N$, which has a kinetic term
$\frac{9}{4}(\partial{N})^2$. The normalization can be fixed by
making the change $N\rightarrow \frac{\sqrt{2}}{3}N$.  We
will fix the normalization later when we go to the unitary gauge and redefine the physical Higgs fields.

We have to know which combinations of these scalars are eaten by massive 
gauge bosons in
order to go to the unitary gauge. This can be done by
investigating the gauge-Higgs mixing terms arising from the covariant
kinetic terms of $H$ and $\hat{H}$. We require all gauge-Higgs
mixing terms vanish after the redefinition of the Higgs fields.
The following re-parametrization corresponds to correct unitary gauge choice 
and are canonically normalized:
\begin{eqnarray}
\label{unitary}
    \begin{array}{ll}
  N~ \rightarrow  \frac{\sqrt{2}\hat{f}}{F(\cos x+2\frac{\sin
x}{x})}\phi^0, & 
~\hat{N}~ \rightarrow  -\frac{\sqrt{2}f\cos x}{3F}\phi^0, \\
  h_1\rightarrow  0, & 
~h_2 \rightarrow  
\frac{v+h}{\sqrt{2}}-i\frac{x\hat{f}}{\sqrt{2}F(\cos x+2\frac{\sin
x}{x})}\phi^0, \\
  C~ \rightarrow  -\frac{x\hat{f}}{F\sin x} \phi^+, & 
~\hat{C}~ \rightarrow  \frac{{f\cos x}}{F} \phi^+. \\
\end{array}
\end{eqnarray}
In these
expressions, we define $F=\sqrt{f^2\cos^2x+\hat{f}^2}$ and
$x=\frac{v}{\sqrt{2}f}$.

\section{Mass formulas and mixing angles}
\label{app:masses}

For completeness, we present the exact expressions of the masses and
mixing matrices for both the gauge and the top sector.

The masses for the massive gauge bosons are 
\begin{eqnarray}\label{numbermasses}
    m^2_{W} &=& \frac{1}{2}g_2^2 f^2 \sin^2x,\\
    m^2_{W_H} &=& \frac{1}{2}g_2^2 (\hat{f}^2+f^2
\cos^2x),\\
    m_{Z}^2 &=&
\frac{g_2^2+g_Y^2}{g_2^2}m_W^2
\frac{2m^2_{W_H}}{m^2_{W_H}+m^2_{W}
+\sqrt{(m^2_{W_H}-m^2_{W})^2
+4\frac{g_1^4}{(g_1^2+g_2^2)^2}m^2_{W_H}m^2_{W}}},\\
    m_{Z_H}^2 &=& \frac{g_1^2+g_2^2}{g_2^2}(m^2_{W_H}+m^2_{W})
-m_{Z}^2,
\end{eqnarray}

The mixing matrix $U$ between the neutral gauge bosons defined in
Eq.~(\ref{eq:mixing}) has the form
\begin{eqnarray}
    U &=& \left(%
\begin{array}{ccc}
  \frac{m^2_{W_H}}{\sqrt{N^+}(m^2_{Z_H}-m^2_{W_H})} 
& \frac{m^2_{W}}{\sqrt{N^+}(m^2_{Z_H}-m^2_{W})} 
& -\frac{g_2}{\sqrt{N^+}g_1} \\
  -\frac{m^2_{W_H}}{\sqrt{N^-}(m^2_{W_H}-m^2_{Z})} 
& \frac{m^2_{W}}{\sqrt{N^-}(m^2_{Z}-m^2_{W})} 
& -\frac{g_2}{\sqrt{N^-}g_1} \\
  \frac{g_1}{\sqrt{2g_1^2+g_2^2}} 
& \frac{g_1}{\sqrt{2g_1^2+g_2^2}} 
& \frac{g_2}{\sqrt{2g_1^2+g_2^2}} \\
\end{array}%
\right).
\end{eqnarray}
$U$ is an unitary matrix with $N^{\pm}$ being the normalization
factors.

The masses  for the light and heavy top quarks are
\begin{eqnarray}
    m_t^2 &=& \frac{1}{2}(M^2+y^2f^2-N_t),\\
    m_{T}^2 &=& \frac{1}{2}(M^2+y^2f^2+N_t),
\end{eqnarray}
where $N_t = \sqrt{(y^2f^2+M^2)^2-y^4f^4\sin^2 2x}$. 

The mixing angles $\alpha_L$ and $\alpha_R$ between top quarks defined in
Eq.~(\ref{eq:topmix}) are
%\begin{eqnarray}
%\tan 2\alpha_L=2\frac{Myf}{y^2f^2\cos 2x+M^2}\sin x\nonumber\\
%\tan 2\alpha_R=2\frac{Myf}{y^2f^2\cos 2x-M^2}\cos x
%\end{eqnarray}%
%
%or
\begin{eqnarray}
\sin \alpha_L&=&\frac{1}{\sqrt{2}}\sqrt{1-(y^2f^2\cos 2x+M^2)/N_t},\\
\sin \alpha_R&=&\frac{1}{\sqrt{2}}\sqrt{1-(y^2f^2\cos 2x-M^2)/N_t}.
\end{eqnarray}

The  field dependent squared masses of the gauge bosons and top quarks
are needed for the calculation of the CW potential.   The masses for 
the charged gauge bosons and top quarks are:
\begin{eqnarray}
    m^2_{W} &= &
\frac{1}{2}g_2^2(H_L^{\dagger}H_L+\hat{H}_L^{\dagger}\hat{H}_L),
\label{MWmass}\\
    m^2_{W_H} &= &
\frac{1}{2}g_2^2(H_R^{\dagger}H_R+\hat{H}_R^{\dagger}\hat{H}_R),
\label{MWHmass}\\
    m_{t}^2 &= &
\frac{1}{2}(M^2+f^2 y^2-\sqrt{(M^2+f^2 y^2)^2-4y^4|H_L|^2(f^2-|H_L|^2)},
\label{Mtmass}\\
    m_{T}^2 &= &
\frac{1}{2}(M^2+f^2 y^2+\sqrt{(M^2+f^2
y^2)^2-4y^4|H_L|^2(f^2-|H_L|^2)},
\label{MTmass}
\end{eqnarray}
where $H_{L(R)}$ is the upper (lower) two components of the Higgs $H$ in 
Eq.~(\ref{GBrep}), and similarly for $\hat{H}_{L(R)}$.

For the squared masses of
the neutral gauge bosons $Z$, $Z_H$ and $\gamma$, we have to solve the following
equation

\begin{eqnarray}
    \lambda^3+a\lambda^2+b\lambda+c=0
\end{eqnarray}
where

\begin{eqnarray}
    a &=& -\frac{g_1^2+g_2^2}{2}(f^2+\hat{f}^2),\\
    b&=&
\frac{2g_1^2+g_2^2}{g_2^2}m_W^2m_{W_H}^2-{\cal P}_L-{\cal P}_R,\\
    c&=& P_Rm_W^2+P_Lm_{W_H}^2,\\
    {\cal P}_L &=&
(g_1g_2)^2\left[|H_L^{\dagger}\hat{H}_L|^2-(H_L^{\dagger}H_L)
(\hat{H}_L^{\dagger}\hat{H}_L)\right],\\
    {\cal P}_R &=&
(g_1g_2)^2\left[|H_R^{\dagger}\hat{H}_R|^2
-(H_R^{\dagger}H_R)(\hat{H}_R^{\dagger}\hat{H}_R)\right].
\end{eqnarray}
Note that $m_W^2$ and $m_{W_H}^2$
appear in the equations above are both field dependent.

All the physical Higgses get masses from both the soft left-right 
symmetry breaking $\mu$-terms, and the
one-loop radiative corrections.   Here we list the 
masses for various Higgses:
\begin{eqnarray}
\label{eq:mass}
    m_{\phi^0}^2&=&\frac{\mu_r^2}{F^2}f\hat{f}
\left[\frac{\hat{f}^2(\cos x+
\frac{\sin x}{x}(3+x^2))}{f^2(\cos x+\frac{\sin x}{x})^2}
+2\cos x+\frac{f^2\cos^2x(1+\cos
    x)}{2\hat{f}^2}\right], \\
    m_{\phi^\pm}^2&=& \frac{3}{16\pi^2}\frac{g_1^2m_{W_H}^2}{m_{Z_H}^2-m_Z^2}\left[(\frac{m_W^2}{m_{Z_H}^2}-1){\cal Z}(m_{Z_H})-(\frac{m_W^2}{m_Z^2}-1){\cal Z}(m_{Z})\right]\nonumber\\
                &&+\frac{\mu_r^2}{F^2}f\hat{f}
\left[\frac{\hat{f}^2x}{f^2\sin x}+2\cos x+
\frac{f^2\cos^3x}{\hat{f}^2}\right],\\
    m_{\hat{h}_2}^2 &=& \frac{3}{16\pi^2}\left[
\frac{m_W^2}{x^2f^2}({\cal Z}(m_W)-{\cal Z}(m_{W_H}))
+\frac{2g_1^2+g_2^2}{4}\frac{m_{W_H}^2
-m_W^2}{m_{Z_H}^2-m_Z^2}({\cal Z}(m_Z)-{\cal Z}(m_{Z_H}))\right]\nonumber \\                &&+\mu_r^2\frac{f}{\hat{f}}\cos x+\hat{\mu}^2,\\
    m_{\hat{h}_1}^2 &=&{m}^2_{\hat{h}_2}
+\frac{3}{16\pi^2}\frac{g_1^2m_W^2}{m_{Z_H}^2-m_Z^2}
\left[(\frac{m_{W_H}^2}{m_{Z_H}^2}-1){\cal
Z}(m_{Z_H})-(\frac{m_{W_H}^2}{m_{Z}^2}-1){\cal Z}(m_{Z})\right].
\end{eqnarray}
where
\begin{eqnarray}
    {\cal Z}(x) &=& -x^2(\ln\frac{\Lambda^2}{x^2}+1).
\end{eqnarray}
We omit the exact mass formula for the SM Higgs since we obtain it 
from the numerical calculation when minimizing the CW potential.

\section{Feynman rules for interactions}
\label{app:LHinteraction}
In this section, we listed the  new vertices which are relevant to
collider physics at the LHC but are not present in the SM.  
The interactions are obtained via expanding the non-linear Higgs 
fields in Eq.~(\ref{GBrep}) up to the fifth order and keeping the leading order terms in 
interactions.

%In most cases, 
%we expand the interactions in order of $v/f$, and keep only the leading
%contributions.

In Table~\ref{tab:gaugehiggs1}, we listed the interactions from
covariant Higgs kinetic term ${\cal L}_H$. Those include
(i) gauge boson-scalar-scalar
interactions, (ii) gauge boson-gauge boson-scalar interactions
(iii) tri-scalar interactions.
For gauge boson-scalar-scalar interactions that are not Hermitian, 
the complex conjugate terms can be obtained by flipping the sign of the real part of 
the coefficients, while keep the imaginary part unchanged.  
In Table~\ref{tab:gaugehiggs2}, we listed 
gauge boson-gauge boson-scalar-scalar interactions.
For terms that are not Hermitian, the complex conjugate terms can be obtained  by
taking the complex conjugation of the coefficients.

There are nonrenormalizable vertices which are 
not listed here but are included in the numerical calculations, 
for example, (i) gauge
boson-scalar-scalar-scalar interactions, and (ii) 
scalar four point interactions. These vertices are of the 
order of $p/f$ and $(p/f)^2$, for $p$ being the momentum of particles.

There are also vertices from the one-loop CW potential. These
vertices contain three- and four- point scalar self-interactions. These
vertices are important compared to the similar interactions from the
kinetic term only at low energy since they are suppressed by loop
factor $\frac{1}{16\pi^2}$ while the latter is proportional to the particle
momentum. In our numerical calculations, the Higgs self-interactions
from CW potentials are also included.  The contribution from those interactions
are usually small.

\begin{table}[tbh]
\begin{tabular}{|l|l|}
  \hline
  % after \\: \hline or \cline{col1-col2} \cline{col3-col4} ...
%$h ~h ~h                $ & $   -3 e MH^2/(2 m_{W} s_w)                                                     $ \\

$\hat{h}_1^{\dagger}~\hat{h}_2 ~W^+_{\mu}       $ & $   -e( p1-p2 )_{\mu}/(\sqrt{2} s_w)          $ \\
%$\hat{h}_2^{\dagger}~ \hat{h}_1 ~W^-_{\mu}       $ & $   -e( p1-p2 )_{\mu}/(\sqrt{2} s_w)          $ \\
$\hat{h}_1^{\dagger}~ \hat{h}_1 ~Z_{\mu}       $ & $   -e(c_w^2-s_w^2)( p1-p2 )_{\mu}/(2 c_w s_w)   $ \\
$\hat{h}_2^{\dagger}~ \hat{h}_2 ~Z_{\mu}        $ & $   ~~e( p1-p2 )_{\mu}/(2 c_w s_w)               $ \\
$\phi^- ~\phi^+ ~ A_{\mu}        $ & $   -e( p1-p2 )_{\mu}                                                 $ \\
$\phi^- ~\phi^+ ~ Z_{\mu}        $ & $   ~~e ( p1-p2 )_{\mu}s_w/c_w                                       $ \\
$h   ~\phi^0 ~Z_{\mu}        $ & $   i ~e~x~p3_{\mu}/(6 c_w s_w)                                      $ \\
$h   ~\phi^0 ~Z_{H\mu}      $ & $   i ~e~x ( (14-17 s_w^2)p2_{\mu}- (4- s_w^2)p1_{\mu} )/(18 s_w c_w c2_w)     $ \\
$\phi^- ~\phi^+  ~Z_{H\mu}      $ & $   -e(1-3 s_w^2)( p1-p2 )_{\mu}/(2 s_w c_w c2_w)                             $ \\
$\hat{h}_1^{\dagger} ~\hat{h}_1 ~Z_{H\mu}       $ & $   ~~e ( p1-p2 )_{\mu}s_w/(2 c_w c2_w)                   $ \\
$\hat{h}_2^{\dagger} ~\hat{h}_2 ~Z_{H\mu}       $ & $   ~~e ( p1-p2 )_{\mu}s_w/(2 c_w c2_w)                   $ \\
$\phi^- ~\phi^0 ~W_{H\mu}^+       $ & $   -e(2 p2-p1)_{\mu}/(3 s_w)                                             $ \\
%$\phi^+ ~\phi^0 ~W_{H\mu}^-       $ & $   ~~e(3 p2-p1)_{\mu}/(4 s_w)                                              $ \\
%$h ~\phi^+ ~ W_{H\mu}^-       $ & $   i~e ~x (2 p2-p1)_{\mu}/(3 s_w)                                          $ \\
$h ~\phi^-  ~ W_{H\mu}^+       $ & $   i~ e~x (2 p2-p1)_{\mu}/(3 s_w)                                          $ \\
$\hat{h}_1^{\dagger} ~\hat{h}_1 ~A_{\mu}      $ & $   -e( p1-p2 )_{\mu}                          $ \\
%\hline
%$h-h-h-W$&\\
\hline
$h~Z_{\mu} ~Z_{\nu}        $ & $   ~~e m_{W}g_{\mu\nu}/(c_w^2 s_w)    $ \\
$h~Z_{\mu} ~Z_{H\nu}       $ & $   e^2f~ x ~g_{\mu\nu}/(\sqrt{2} c_w^2 c2_w)  $ \\
$h~Z_{H\mu}~ Z_{H\nu}       $ & $   -e^2f~ x~ g_{\mu\nu}/(\sqrt{2} c_w^2 s_w^2)   $ \\
$h~W^+_{\mu}~ W^-_{\nu}       $ & $   ~~e m_{W}g_{\mu\nu}/s_w                                    $ \\
$h~W_{H\mu}^+~ W_{H\nu}^-       $ & $   -e^2f~ x~ g_{\mu\nu}/(\sqrt{2} s_w^2)                     $ \\
\hline
$\phi^- ~\phi^+ ~h            $ & $   ~~x(p3\cdot p3+2 p1\cdot p2)/(3 \sqrt{2} f)           $ \\
$\phi^- ~\phi^+ ~\phi^0         $ & $   i~p3\cdot(p2-p1)/(3 \sqrt{2} f)                  $ \\
$h ~\phi^0 ~\phi^0          $ & $   ~~x( 30 p2\cdot p3+11 p1\cdot p1)/(27 \sqrt{2} f)           $ \\
$\hat{h}_1^{\dagger} ~\hat{h}_1 ~\phi^0    $ & $   i~ f~ p3\cdot(p1-p2)/(3 \sqrt{2} \hat{f}^2)     $ \\
$\hat{h}_2^{\dagger} ~\hat{h}_2 ~\phi^0    $ & $   i ~f ~p3\cdot(p1-p2)/(3 \sqrt{2} \hat{f}^2)       $ \\
\hline
\end{tabular}
\caption{New scalar self interactions and scalar-gauge boson interactions from the
covariant kinetic terms of the Higgses.  $p1$, $p2$ and $p3$
refer to the incoming momentum of the first, second and third particle, respectively.}
\label{tab:gaugehiggs1}
\end{table}

\begin{table}[tbh]
\begin{tabular}{|l|l|l|l|}
  \hline

$h ~h ~W^+_{\mu} ~W^-_{\nu} $ & $   ~~e^2g_{\mu\nu}/(2 s_w^2)$& $h
~h    ~W^+_{H\mu} ~W^-_{H\nu} $ & $ -e^2g_{\mu\nu}/(2s_w^2)$\\
$h ~h ~Z_{\mu} ~Z_{\nu}     $ & $   ~~e^2g_{\mu\nu}/(2 c_w^2 s_w^2)
$&$h    ~h    ~Z_{H\mu} ~Z_{H\nu} $ &
$-e^2g_{\mu\nu}/(2c_w^2s_w^2)  $\\
$h    ~h ~Z_{\mu} ~Z_{H\nu} $ & $~~e^2g_{\mu\nu}/(2 c_w^2 c2_w) $& &\\
$\phi^0 ~\phi^0   ~W^+_{\mu} ~W^-_{\nu}              $ & $ -e^2
x^2g_{\mu\nu}/(54 s_w^2) $&$\phi^0   ~\phi^0 ~W^+_{H\mu}
~W^-_{H\nu} $ & $ ~~e^2 x^2g_{\mu\nu}/(54 s_w^2)$\\
$\phi^0   ~\phi^0 ~Z_{\mu}   ~Z_{\nu} $ & $
-e^2x^2g_{\mu\nu}/(54s_w^2c_w^2)  $&$\phi^0   ~\phi^0  ~Z_{H\mu}
~Z_{H\nu} $ & $ ~~e^2 x^2g_{\mu\nu}/(54 s_w^2c_w^2)  $\\
$\phi^0 ~\phi^0 ~Z_{\mu}    ~Z_{H\nu} $ & $ -e^2 x^2g_{\mu\nu}/(54
c_w^2c2_w) $&&\\
$\phi^-   ~\phi^0    ~W^+_{H\mu} ~A_{\nu} $ & $-e^2 g_{\mu\nu}/(3
s_w) $&$\phi^-  ~h     ~W^+_{H\mu} ~A_{\nu} $ & $-2ie^2  xg_{\mu\nu}/(3s_w)  $\\
$\phi^-   ~\phi^0 W^+_{H\mu} ~Z_{\nu} $ & $~~e^2g_{\mu\nu}/(3c_w)
$&$\phi^- ~h  ~W^+_{H\mu} ~Z_{\nu} $ & $~~2i e^2 x
g_{\mu\nu}/(3c_w)  $\\
$\phi^- ~\phi^0 W^+_{H\mu} ~Z_{H\nu} $ & $ ~~e^2 g_{\mu\nu}/(3c_w
c2_w) $&$\phi^-   ~h    ~W^+_{H\mu}  ~Z_{H\nu} $ & $~~2i e^2xg_{\mu\nu}/(3c_w c2_w)  $\\
$\phi^- ~\phi^+ ~Z_{\mu} ~Z_{\nu}     $ & $~~ 2 e^2
s_w^2g_{\mu\nu}/c_w^2  $&
$\phi^- ~\phi^+ ~A_{\mu} ~A_{\nu} $ & $~~2e^2 g_{\mu\nu} $
\\
$\phi^- ~\phi^+ ~Z_{\mu} ~Z_{H\nu} $ & $
-e^2(3c_w^2-2)g_{\mu\nu}/( c2_w c_w^2)  $&
$\phi^- ~\phi^+ ~Z_{\mu} ~A_{\nu}  $ & $-2 e^2 s_wg_{\mu\nu}/c_w  $\\
$\phi^- ~\phi^+    ~Z_{H\mu}   ~Z_{H\nu}$ & $-2
e^2g_{\mu\nu}/c_w^2 $&
$\phi^- ~\phi^+    ~Z_{H\mu}   ~A_{\nu} $ & $~~e^2(3c_w^2-2)g_{\mu\nu}/( c2_w c_w s_w)  $\\
$\phi^- ~\phi^+ ~W^+_{\mu} ~W^-_{\nu} $ & $ -e^2 x^2 g_{\mu\nu}/(6 s_w^2)  $&
$\phi^- ~\phi^+ ~W^+_{H\mu} ~W^-_{H\nu} $ & $ ~~e^2 x^2 g_{\mu\nu}/(6 s_w^2)  $\\
$\hat{h}_1^{\dagger} ~\hat{h}_1 ~W^+_{\mu} ~W^-_{\nu}   $ & $
~~e^2g_{\mu\nu}/(2s_w^2)  $& $\hat{h}_2^{\dagger} ~\hat{h}_2
~W^+_{\mu} ~W^-_{\nu}   $ & $~~ e^2g_{\mu\nu}/(2s_w^2)  $\\
$\hat{h}_1^{\dagger} ~\hat{h}_1  ~W^+_{H\mu} ~W^-_{H\nu} $ &
$-e^2g_{\mu\nu}/(2 s_w^2)  $& $\hat{h}_2^{\dagger} ~\hat{h}_2
~W^+_{H\mu} ~W^-_{H\nu} $ & $ -e^2g_{\mu\nu}/(2 s_w^2)  $\\
$\hat{h}_1^{\dagger} ~\hat{h}_1 ~Z_{\mu}    ~Z_{\nu} $ & $~~e^2
c2_w^4g_{\mu\nu} /(2 c_w^2 s_w^2) $&$\hat{h}_2^{\dagger} ~\hat{h}_2  ~Z_{\mu}    ~Z_{\nu} $ & $~~e^2g_{\mu\nu}/(2s_w^2c_w^2)  $\\
$\hat{h}_1^{\dagger} ~\hat{h}_1  ~Z_{\mu} ~Z_{H\nu} $ &
$-e^2c2_wg_{\mu\nu}/(2c_w^2) $&$\hat{h}_2^{\dagger} ~\hat{h}_2
~Z_{\mu} ~Z_{H\nu} $ &~~$e^2g_{\mu\nu}/(2c_w^2c2_w)  $\\
$\hat{h}_1^{\dagger} ~\hat{h}_1   ~Z_{H\mu}  ~Z_{H\nu} $ &
$-e^2g_{\mu\nu}/(2s_w^2c_w^2)  $&$\hat{h}_2^{\dagger} ~\hat{h}_2   ~Z_{H\mu} ~Z_{H\nu} $ & $-e^2g_{\mu\nu}/(2s_w^2c_w^2)  $\\
$\hat{h}_1^{\dagger} ~\hat{h}_1 ~Z_{\mu} ~A_{\nu} $ & $~~e^2c2_w^2g_{\mu\nu}/(c_ws_w)  $&$\hat{h}_1^{\dagger} ~\hat{h}_1  ~Z_{H\mu} ~A_{\nu} $ & $-e^2s_wg_{\mu\nu}/(c_wc2_w)  $\\
$\hat{h}_1^{\dagger} ~\hat{h}_1  ~A_{\mu} ~A_{\nu} $ & $~~2e^2
g_{\mu\nu} $& &\\
$\hat{h}_1^{\dagger}  ~\hat{h}_2 W^+_{\mu} ~Z_{\mu} $ &
$-e^2g_{\mu\nu}/(\sqrt{2}c_w)  $&
$\hat{h}_1^{\dagger}  ~\hat{h}_2 W^+_{\mu} ~Z_{H\nu} $ & $-e^2g_{\mu\nu}/(\sqrt{2}c_wc2_w)  $\\
$\hat{h}_1^{\dagger}  ~\hat{h}_2 W^+_{\mu} ~A_{\nu} $ &
$~~e^2g_{\mu\nu}/(\sqrt{2} s_w)  $&
&\\

%\hline
%$h ~h ~h ~h     $ & $   -3 e^2 MH^2/(4 m_{W}^2 s_w^2) $
\hline
\end{tabular}
\caption{Four point gauge boson-gauge boson-scalar-scalar 
interactions from the covariant kinetic terms of the
scalar.} 
\label{tab:gaugehiggs2}
\end{table}

\begin{table}[tbh]
\begin{tabular}{|l|l|}
  \hline
  % after \\: \hline or \cline{col1-col2} \cline{col3-col4} ...
$A_{\mu}~W^+_{\nu} ~W^-_{\eta}       $ & $   ~~e ~F_{\mu\nu\eta}   $ \\
$Z_{\mu}~W^+_{\nu}    ~W^-_{\eta}          $ & $   ~~e ~F_{\mu\nu\eta}~c_w /s_w       $ \\
$Z_{H\mu}~W^+_{\nu}    ~W^-_{\eta}         $ & $   ~~e  ~F_{\mu\nu\eta}~(c2_w s_w /c_w^3)(m_W^2/ m_{W_H}^2)          $ \\
$A_{\mu}~W_{H\nu}^+   ~ W_{H\eta}^-         $ & $   ~~e   ~F_{\mu\nu\eta}         $ \\
$Z_{\mu}~W_{H\nu}^+    ~W_{H\eta}^-          $ & $   -e    ~F_{\mu\nu\eta}~s_w/c_w             $ \\
$Z_{H\mu}~W_{H\nu}^+    ~W_{H\eta}^-          $ & $   ~~e    ~F_{\mu\nu\eta}~c2_w/(s_w c_w)    $ \\
\hline
$W^+_{\mu}    ~W^+_{\nu}    ~W^-_{\eta}    ~W^-_{\rho} $ & $   ~~e^2                                         ~G_{\mu\nu\eta\rho}/s_w^2         $ \\
$W_{H\mu}^+    ~W_{H\nu}^+    ~W_{H\eta}^-    ~W_{H\rho}^- $ & $   ~~e^2                                     ~G_{\mu\nu\eta\rho}/s_w^2        $ \\
$A_{\mu}     ~A_{\nu}     ~W^+_{\eta}    ~W^-_{\rho} $ & $   -e^2                                          ~G_{\mu\nu\eta\rho}                                           $ \\
$Z_{\mu}~Z_{\nu} ~W^+_{\eta}    ~W^-_{\rho}     $ & $   -e^2                                               ~G_{\mu\nu\eta\rho} ~c_w^2 /s_w^2           $ \\
$A_{\mu}     ~Z_{\nu}~W^+_{\eta}    ~W^-_{\rho}  $ & $   -e^2                                             ~G_{\mu\nu\eta\rho}~c_w /s_w              $ \\
$A_{\mu}     ~Z_{H\nu} ~W_{\eta}^+    ~W_{\rho}^-     $ & $   -e^2        ~G_{\mu\nu\eta\rho}~(s_w c2_w /c_w^3)(m_{W}^2/ m_{W_H}^2)    $ \\
$Z_{\mu}      ~Z_{H\nu}~W_{\eta}^+    ~W_{\rho}^-    $ & $   -e^2             ~G_{\mu\nu\eta\rho}~(c2_w /c_w^2) (m_{W}^2/m_{W_H}^2)         $ \\
$Z_{H\mu}    ~Z_{H\nu} ~W_{\eta}^+    ~W_{\rho}^-  $ & $   -e^2      ~G_{\mu\nu\eta\rho}~(s_w^2 c2_w^2 /c_w^6)(m_{W}^4/ m_{W_H}^4)   $ \\
$A_{\mu}~A_{\nu}~W_{H\eta}^+    ~W_{H\rho}^-    $ & $   -e^2                                               ~G_{\mu\nu\eta\rho}   $ \\
$Z_{\mu}~Z_{\nu}~W_{H\eta}^+    ~W_{H\rho}^-  $ & $   -e^2                                                 ~G_{\mu\nu\eta\rho}~s_w^2/c_w^2        $ \\
$A_{\mu}     ~Z_{\nu}~W_{H\eta}^+    ~W_{H\rho}^-   $ & $   ~~e^2                                           ~G_{\mu\nu\eta\rho}~s_w/c_w                   $ \\
$A_{\mu}     ~Z_{H\nu} ~W_{H\eta}^+    ~W_{H\rho}^-     $ & $   -e^2                          ~G_{\mu\nu\eta\rho}~c2_w/(c_w s_w)   $ \\
$Z_{\mu}     ~Z_{H\nu} ~W_{H\eta}^+    ~W_{H\rho}^-     $ & $   ~~e^2                           ~G_{\mu\nu\eta\rho}~c2_w/c_w^2    $ \\
$  Z_{H\mu}    ~Z_{H\nu}~W_{H\eta}^+    ~W_{H\rho}^-   $ & $   -e^2                    ~G_{\mu\nu\eta\rho}~c2_w^2/(s_w^2 c_w^2)      $ \\
\hline
\end{tabular}
\caption{ Gauge self couplings with the spin structure
$F_{\mu\nu\eta}=(p3 -p2)_{\mu} g_{\nu\eta}+(p1 -p3)_{\nu}
g_{\mu\eta}+(p2 -p1)_{\eta} g_{\mu\nu}$ and $G_{\mu\nu\eta\rho}=2
g_{\mu\nu} g_{\eta\rho}-g_{\mu\eta} g_{\nu\rho}-g_{\mu\rho}
g_{\nu\eta}$. We also use the notations $s_w=\sin\theta_w$,
$c_w=\cos\theta_w$ and $c2_w=\sqrt{\cos2\theta_w}$ of the Weinberg
angle $\theta_w$.}
\label{tab:gaugeself}
\end{table}

The gauge self-couplings between the gauge boson mass
eigenstates can be obtained from the kinetic terms 
for the ${\rm SU}(2)_L$ and ${\rm SU}(2)_R$ gauge bosons,
using  the mixing matrix $U$ for the  neutral gauge bosons
given in Eq.~(\ref{eq:mixing}).
In Table~\ref{tab:gaugeself}, we
listed all the gauge self-interactions.
%
%\begin{eqnarray}
%    {\cal L}_{{\rm SU}(2)-self} &=& -ig_2\left[%
%W_{\mu}^{+}(W_{\nu}^0\overrightarrow{\partial}^{\mu}W^{-\nu})-W_{\mu}^{-}(W_{\nu}^0\overrightarrow{\partial}^{\mu}W^{+\nu})+W_{\nu}^{0}W_{\mu}^-\partial^{\nu}W^{+\mu}
%-W_{\nu}^{0}W_{\mu}^+\partial^{\nu}W^{-\mu}\right]%
%\nonumber\\
%&&-\frac{1}{2}g^2_2[ (W^+\cdot W^-)^2-(W^+\cdot W^+)(W^-\cdot
%W^-)+2(W^0\cdot W^0)(W^+\cdot W^-)\nonumber\\
%&&\;\;\;\;\;\;\;\;\;-2(W^0\cdot W^+)(W^0\cdot W^-)]%
%\end{eqnarray}
%where $A\overrightarrow{\partial}B = A\partial B-B\partial A$. 
%The self interactions between gauge boson mass eigenstates can be
%obtained by using the mixing matrix between neutral gauge bosons
%found in eq.(\ref{eq:mixing}). 

In Table~\ref{tab:higgsfermion} we listed the
Higgs-fermion-fermion interactions.
In Table~\ref{tab:gaugefermion}, we
listed the gauge-fermion-fermion interactions, where we have
ignored the flavor mixing for the charge current. 
Note that for
term which is not Hermitian, the Hermitian conjugate term must also be
added. This can be done by taking the complex conjugate of the
coefficient and, for the Higgs-fermion-fermion interactions,
exchanging $P_L\leftrightarrow P_R$. 
%With the left-right symmetry
%we have defined, the CKM matrix for the right handed fermions
%couple to $W_H^{\pm}$ is same as that of the left handed fermions
%couple to $W^{\pm}$.

\begin{table}[tbh]
\begin{tabular}{|l|l|}
  \hline
%$h~\bar{q}~q  $ & $          -e m_q/(2 m_{W} s_w)$ \\
%$h~\bar{l}~l  $ & $          -e m_{l}/(2 m_{W} s_w)   $ \\
$h~\bar{t}~t  $ & $          -e m_t C_L C_R/(2 m_{W} s_w) $ \\
$h~\bar{T} ~t   $ & $      - y((C_L S_R+S_L C_R x)  P_L+(C_L S_R x+S_L C_R)  P_R)/\sqrt{2} $ \\
$h~\bar{T} ~T $ & $      -y (S_R S_L-C_L C_R x)/\sqrt{2}    $ \\
$\phi^0~\bar{T} ~t   $ & $   -i y ( S_L C_R  P_L -C_L S_R  P_R )/\sqrt{2}  $ \\
$\phi^0~\bar{T} ~T $ & $   -i yC_L C_R\gamma_5/\sqrt{2}    $ \\
$\phi^0~\bar{t}~t      $ & $   -i yS_R S_L \gamma_5 /\sqrt{2}  $ \\
$\phi^0~\bar{b} ~b     $ & $   ~~i m_b \gamma_5/(\sqrt{2} f)  $ \\
$\phi^0~\bar{u}_{1,2} ~u_{1,2}     $ & $   -i m_{u_i} \gamma_5/(\sqrt{2} f)  $ \\
$\phi^0~\bar{d}_{1,2} ~d_{1,2}     $ & $   ~~i m_{d_i} \gamma_5/(\sqrt{2} f)  $ \\
%$\phi^0~\bar{\nu}_{1,2,3} ~\nu_{1,2,3}     $ & $   -i m_{\nu_i} \gamma_5/(\sqrt{2} f)  $ \\
$\phi^0~\bar{l} ~l     $ & $   ~~i m_{l} \gamma_5/(\sqrt{2} f)  $ \\
$\phi^+~\bar{t} ~b       $ & $  -i (S_R m_b  P_L-y  S_L f P_R)/ f        $ \\
$\phi^+~\bar{T} ~b    $ & $   ~~i (C_R m_b  P_L-y C_L f P_R)/f          $ \\
$\phi^+~\bar{u}_{1,2} ~d_{1,2}    $ & $   ~~i (m_{d_i} P_L-m_{u_i}P_R)/f          $ \\
$\phi^+~\bar{\nu}_l ~l    $ & $   ~~i m_{l} P_L/f          $ \\
\hline
\end{tabular}
\caption{ A summary of the new Higgs-fermion-fermion interactions.
$P_{R,L}=\frac{1}{2}(1\pm\gamma_5)$ are the chirality projection
operators.}
 \label{tab:higgsfermion}
\end{table}

\begin{table}[tbh]
\begin{tabular}{|l|l|}
  \hline
$A_{\mu}~\bar{T}~ T     $ & $            ~~2 e  \gamma_{\mu}/3                                   $ \\
$Z_{\mu}~\bar{t}~t          $ & $            ~~e \gamma_{\mu}          ((3 C_L^2-4 s_w^2) P_L-4 s_w^2  P_R)/(6 c_w s_w)   $ \\
$Z_{\mu}~\bar{T}~T      $ & $            -e \gamma_{\mu}(4 s_w^2 -3 S_L^2  P_L)/(6 c_w s_w)  $ \\
$Z_{\mu}~\bar{T}~t        $ & $            ~~e \gamma_{\mu}(  C_L S_L \hat{f}^2 c_w^2P_L+  f^2 x^2 s_w^2 C_R S_RP_R)/(2 \hat{f}^2 c_w^3 s_w) $ \\
$W^+_{\mu}~\bar{t}~b      $ & $            ~~e \gamma_{\mu} C_LP_L/( \sqrt{2} s_w)                     $ \\
$W^+_{\mu}~\bar{T}~b      $ & $            ~~e \gamma_{\mu} S_LP_L/( \sqrt{2} s_w)                     $ \\
$W_{H\mu}^+~\bar{T}~b    $ & $            ~~e \gamma_{\mu}C_R   P_R/( \sqrt{2} s_w)                       $ \\
$W_{H\mu}^+~\bar{t}~b    $ & $            -e \gamma_{\mu}S_R   P_R/( \sqrt{2} s_w)                       $ \\
$W_{H\mu}^+~\bar{u}_{1,2}~d_{1,2}    $ & $            ~~e \gamma_{\mu}   P_R/( \sqrt{2} s_w)                       $ \\
$W_{H\mu}^+~\bar{\nu}_l~l    $ & $            ~~e \gamma_{\mu}   P_R/( \sqrt{2} s_w)                       $ \\
$Z_{H\mu}~\bar{t}~t        $ & $            -e \gamma_{\mu}      ( (1+3 S_L^2) s_w^2P_L-  (3 c_w^2 S_R^2-4 s_w^2)P_R)/(6 c_w c2_w s_w)  $ \\
$Z_{H\mu}~\bar{T}~t      $ & $            ~~e \gamma_{\mu}(C_L S_L s_w^2P_L  - C_R S_R c_w^2P_R )/(2 s_w c_w c2_w)  $ \\
$Z_{H\mu}~\bar{T}~T   $ & $            -e \gamma_{\mu}(  (3 C_L^2+1) s_w^2P_L-  (3 c_w^2 C_R^2-4 s_w^2)P_R )/(6 c_w c2_w s_w)  $ \\
$Z_{H\mu}~\bar{b}~b        $ & $            -e \gamma_{\mu}( s_w^2P_L+  (3-5 s_w^2) P_R )/(6 s_w c_w c2_w)  $ \\
$Z_{H\mu}~\bar{u}_{1,2}~ u_{1,2}   $ & $    -e \gamma_{\mu} (2 s_w^2  P_L+(1-7 c2_w^2)  P_R )/(12 c_w s_w c2_w)  $ \\
$Z_{H\mu}~\bar{d}_{1,2}~ d_{1,2}   $ & $    -e \gamma_{\mu}( s_w^2+  (3-5 s_w^2) P_R )/(6 s_w c_w c2_w)  $ \\
$Z_{H\mu}~\bar{l}~l        $ & $            ~~e \gamma_{\mu} (2s_w^2 P_L+(1-3 c2_w^2)  P_R)/(4 c_w s_w c2_w) $ \\
%$Z~\bar{\nu}~\nu      $ & $            -e \gamma_{\mu} (m_{W_H}^2  P_L-m_Z^2 s_w^2  P_R)/(2 c_w s_w m_{W_H}^2)      $ \\
$Z_{H\mu}~\bar{\nu}_L~\nu_L    $ & $            ~~e \gamma_{\mu} s_w /(2 c_w c2_w)   P_L                  $ \\
$Z_{H\mu}~\bar{\nu}_R~\nu_R    $ & $            ~~e \gamma_{\mu}c_w   /(2  s_w c2_w)      P_R               $ \\
$Z_{\mu}~\bar{\nu}_R~\nu_R      $ & $            -e \gamma_{\mu} m_Z^2 s_w /(2 c_w m_{W_H}^2)      P_R$ \\

\hline
\end{tabular}

\caption{ A summary of new gauge-fermion-fermion interactions.
$P_{R,L}=\frac{1}{2}(1\pm\gamma_5)$ are the chirality projection
operators. The mixing angles $C_L=\cos\alpha_L$,
$C_R=\cos\alpha_R$ etc. are given in Eq.~(\ref{topmixing}). 
}
\label{tab:gaugefermion}
\end{table}

%\end{appendix}

\end{document}